\documentclass[letterpaper,twocolumn,10pt]{article}
\usepackage{usenix}

\usepackage{amsmath,amssymb}
\usepackage{booktabs}
\usepackage{graphicx}
\usepackage{float}
\usepackage{dblfloatfix}
\usepackage{tikz}
\hypersetup{hidelinks}

\DeclareRobustCommand{\fbFull}{\tikz[baseline=-0.55ex,inner sep=0pt,outer sep=0pt]\filldraw[line width=0.3pt] (0,0) circle (0.63ex);}
\DeclareRobustCommand{\fbHalf}{\tikz[baseline=-0.55ex,inner sep=0pt,outer sep=0pt]{%
  \draw[line width=0.3pt] (0,0) circle (0.63ex);%
  \fill (0,0) -- (90:0.63ex) arc (90:270:0.63ex) -- cycle;}}
\DeclareRobustCommand{\fbEmpty}{\tikz[baseline=-0.55ex,inner sep=0pt,outer sep=0pt]\draw[line width=0.3pt] (0,0) circle (0.63ex);}

\begin{document}

\date{}

\title{\Large \bf FlyBlind: Cross-Slice Timeliness Attacks on UAV Situational Awareness over 5G}

\author{
{\rm Wagner Comin Sonaglio$^*$ \quad Ágney Lopes Roth Ferraz$^*$ \quad Sidnei Barbieri$^*$ \quad André Elias Melo$^*$}\\
{\rm Guevara Noubir$^\dagger$ \quad Lourenço Alves Pereira Júnior$^*$}\\
\\
{\rm $^*$Aeronautics Institute of Technology}\\
{\rm Email: sonaglio@ita.br, roth@ita.br, sidneisb@ita.br, andre.melo@ita.br, ljr@ita.br}\\
{\rm $^\dagger$Northeastern University}\\
{\rm Email: g.noubir@northeastern.edu}
}

\maketitle

\begin{abstract}
Beyond Visual Line of Sight (BVLOS) Uncrewed Aerial Systems (UAS) operating over 5G Standalone (SA) networks use a shared User Plane for both command-and-control (C2) data and video feedback. Operators assess link quality through latency and availability, relying on soft isolation between network slices. However, the risk that an authorized co-tenant could make the Ground Control Station (GCS) state outdated without disrupting the connection remains underexplored. This work introduces FlyBlind, a timeliness attack in which an authorized co-tenant on a neighboring slice maintains legitimate uplink demand, causing state aging at the GCS without a rogue gNB or direct interference with C2 traffic. Our key insight is that, under soft isolation, sharing idle resources turns authorized competition for grants into state aging that conventional link monitors fail to detect. We formalize this effect, termed Silent State Staleness, as a falsifiable false-healthy predicate. On a dedicated testbed, telemetry age at the GCS saturates at approximately 12 seconds, with discrepancies of tens of meters between the GCS position estimate and ground truth, while the one-way delay (OWD) p99 remains in the tens of milliseconds, availability exceeds 99.9\%, and the vehicle keeps operating in GUIDED mode without triggering failsafe mechanisms. These findings indicate that, in deployments with asymmetric uplink enforcement, verifying state freshness at the destination is essential rather than relying solely on link health.
\end{abstract}

\begin{figure}[t]
\centering
\includegraphics[width=\columnwidth]{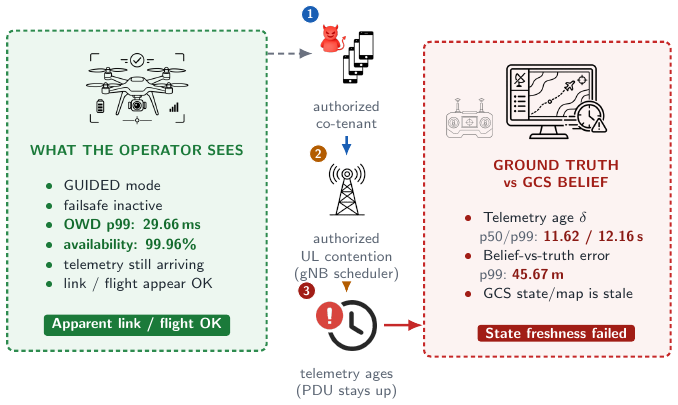}
\caption{Silent State Staleness under FlyBlind. Steps 1--3 are the attack mechanism.}
\label{fig:teaser}
\end{figure}

\section{Introduction}
\label{sec:intro}

Cellular connectivity has become a cornerstone of Beyond Visual Line of Sight (BVLOS) drone operations, enabling wide-area missions under broad coverage, mobility, and standardized network support~\cite{fotouhi_survey_uav_cellular_2019,geraci_what_2022,abdalla_communications_standards_unmanned_2021}.
Operators increasingly rely on 5G Standalone (SA) networks to carry command-and-control (C2), telemetry, and auxiliary flight video feedback (video used to aid Uncrewed Aerial Vehicle (UAV) control~\cite{3gpp_ts_22125}), multiplexed on the same User Plane or on neighboring sessions identified by Single Network Slice Selection Assistance Information (S-NSSAI).
3GPP has standardized Uncrewed Aerial System (UAS) support over cellular networks~\cite{3gpp_ts_22125}.
Network slicing enables differentiated and isolated services on shared radio infrastructure~\cite{elayoubi_5g_ran_slicing_2019}.

We consider a Ground Control Station (GCS) supervising a UAS over a 5G SA network, with a vehicle User Equipment (UE) carrying C2 and video feedback on the User Plane, a shared gNodeB (gNB), and one or more co-tenant UEs on a distinct S-NSSAI attached to the same cell.
The system must preserve the freshness with which the GCS observes vehicle state, formalized by Age of Information (AoI), which is fundamental when a system monitors a physical process through delayed updates~\cite{yates_age_information_introduction_2021,tripathi_wiswarm_age_information_2022}.
The system's benign behavior is a GCS whose belief tracks the real vehicle within a bounded delay; we refer to authorized uplink traffic sent for its own purpose as benign, and to authorized uplink traffic sustained to age another tenant's state as adversarial.

When soft isolation borrows idle radio resources across slices, this enables an authorized co-tenant to sustain legitimate uplink on a neighbor slice and age state at the GCS, without a rogue gNB and without touching the C2 flow, by turning competition for grants and Physical Resource Blocks (PRBs) into telemetry delay. The asymmetry matters because telemetry that carries state to the operator rises on the uplink, while the downlink that carries commands stays healthy.
The operator sees healthy link indicators and a vehicle that appears on course, while the position on the map has already fallen behind and the real aircraft has advanced by tens of meters.

Silent State Staleness poses significant risks in real BVLOS deployments, including misinformed navigation decisions, degraded deconfliction between the operator and the airspace, and supervision carried out on a state that no longer holds. 
Latency, availability, and integrity at the link are monitored in industry practice and in standardized 5G KPI frameworks~\mbox{\cite{honeywelltestreport263do377c2linktest_us_2023,3gpp_ts_28554}}.
Destination freshness for remote situational awareness nevertheless remains under-specified.
Despite its growing importance, state freshness under an authorized adversary remains largely unexplored.

Research on slicing and UAS over cellular networks splits into benign performance studies and security with stronger adversaries.
Benign studies show that proportional fair (PF) scheduling can shift telemetry and apparent trajectory relative to real flight, while slicing improves the behavior~\cite{baguer_enabling_2024}, and that Radio Access Network (RAN) slice guarantees still leave application-level service assurance open~\cite{balasingam_application_level_service_2024}.
Security studies assume adversaries outside the subscriber boundary, such as a rogue gNB manipulating slice allocation~\cite{xu_integrity_under_siege_2025}, physical-layer side channels and jamming~\cite{ashik_dosq_cross_layer_2026}, or User Plane Function (UPF) contention~\cite{moreira_noisy_neighbor_influence_2026}.
In closing, Baguer et al.\ call for BVLoS pilot-side metrics, among them ``new BVLoS-specific QoE metrics that measure, from the pilot perspective, the relative difference between FPV video and received telemetry''~\cite{baguer_enabling_2024}.

Both lines leave destination freshness unmeasured while the adversary stays inside the subscriber boundary.
Our key insight is that sharing of idle resources under soft isolation, foreseen in TS~28.541~\cite{3gpp_ts_28541} for efficiency in dynamic slicing, can convert into state aging for a freshness-sensitive tenant without conventionally monitored latency, availability, or integrity revealing that effect.

In this work, we present FlyBlind, a timeliness attack that materializes this insight under uplink (UL) cross-slice contention between S-NSSAIs with incomplete enforcement~\cite{elayoubi_5g_ran_slicing_2019,vittal_preventing_cross_network_2023} in a Software-in-the-Loop (SITL) testbed with ArduPilot~\cite{baldi_ardupilot_based_adaptive_2022}.
An authorized co-tenant injects schedulable uplink demand on a neighbor slice to create divergence between the GCS state estimate and the vehicle's physical state, which we characterize as Silent State Staleness and formalize in a falsifiable false-healthy predicate (Eq.~\eqref{eq:sss}).
On a 5G SA testbed with RAN slicing and a closed control loop, across adversarial doses, video feedback rates, and radio bandwidths, telemetry age at the GCS saturates around 12~s. The displayed position diverges by tens of meters. At the same time, one-way delay (OWD) p99 stays in the tens of milliseconds, availability stays above 99.9\%, and the vehicle keeps flying in GUIDED mode without triggering a failsafe.
Figure~\ref{fig:teaser} shows the predicate in operational form: conventional monitors stay green while freshness and belief-vs-truth divergence increase. Hence, the operator reads the link as OK while state at the GCS is stale~\cite{huang_gray_failure_2017}.

The attacked axis is timeliness, and not availability.
The PDU Session remains, packets flow, and the vehicle flies while the GCS state becomes outdated.
In contrast, denial-of-service (DoS) in the classical taxonomy consumes a key resource until it becomes unavailable~\cite{mirkovic_taxonomy_ddos_attack_2004}, an axis distinct from ours.
Unlike Xu et al., who assume a rogue gNB capable of manipulating slice allocation~\cite{xu_integrity_under_siege_2025}, and DoSQ, which assumes side channels and jamming on the physical layer or on Downlink Control Information (DCI)~\cite{ashik_dosq_cross_layer_2026}, FlyBlind requires only an authorized subscriber with legitimate UL traffic on a neighbor slice.

In summary, our contributions are as follows.
\begin{enumerate}
\item \textbf{Freshness as an attacked property with a falsifiable false-healthy predicate.} We characterize state freshness as a property under attack in UAS C2 over 5G, beyond a benign metric~\cite{yates_age_information_introduction_2021,tripathi_wiswarm_age_information_2022,baguer_enabling_2024}. We formalize and validate a falsifiable false-healthy predicate (Eq.~\eqref{eq:sss}, with $H_{\mathrm{net}}{=}1$, $H_{\mathrm{veh}}{=}1$, $F_{\mathrm{GCS}}{=}0$) that separates link KPIs from state freshness at the GCS (Tab.~\ref{tab:false-healthy}, Eq.~\eqref{eq:afresh})~\cite{huang_gray_failure_2017}.

\item \textbf{Authorized adversary with cyber-physical consequence.} We show that an authorized co-tenant, without rogue gNB, without jamming, and without touching the victim C2 flow, induces Silent State Staleness with measurable divergence in GCS belief and stable physical flight (Sec.~\ref{sec:attack}, Tab.~\ref{tab:false-healthy}), in contrast with Xu et al.~\cite{xu_integrity_under_siege_2025} and DoSQ~\cite{ashik_dosq_cross_layer_2026}.

\item \textbf{Freshness-aware observability for C2 over 5G.} We instrument delivery age ($\delta_i$) and \texttt{belief\_vs\_truth} as security observables at the destination and show that monitored latency, availability, and integrity remain green. At the same time, the operational property fails, answering the pilot-side need stated by Baguer et al.~\cite{baguer_enabling_2024}.

\item \textbf{Causal attribution and the cost of isolation.} We attribute the aging through convergent controls that rule out static internal queue, scheduling cap, and clock offset as dominant causes (Sec.~\ref{sec:lcg-hol}, Tab.~\ref{tab:pre-attack-drain}) and show that an uplink floor on the victim slice restores freshness (B20--B80, Tab.~\ref{tab:mitigations-map}) with the lever in \texttt{rRMPolicyMinRatio}~\cite{3gpp_ts_28541}. At the same time, we characterize the cost of that isolation through ReservationCost and ReservationWaste (Tab.~\ref{tab:frontier-sweep}).
\end{enumerate}

This paper is structured as follows.
Section~\ref{sec:background} presents background and motivation; Section~\ref{sec:threat} defines the threat model; Section~\ref{sec:attack} describes the FlyBlind attack; Section~\ref{sec:setup} details the experimental setup; Section~\ref{sec:eval} reports evaluation results; Section~\ref{sec:mitigations} discusses mitigations and the cost of isolation; Section~\ref{sec:discussion} covers discussion and threats to validity; Section~\ref{sec:related} reviews related work; and Section~\ref{sec:conclusion} concludes.

\section{Background and Motivation}
\label{sec:background}

\subsection{RAN Slicing and RRMPolicyRatio}

The RAN slicing model described in 3GPP already foresees division and isolation policies based on quotas and priorities.
In the Network Resource Model (TS~28.541)~\cite{3gpp_ts_28541}, the IOC \texttt{RRMPolicyRatio} exposes \texttt{rRMPolicyMinRatio}, \texttt{rRMPolicyDedicatedRatio}, and \texttt{rRMPolicyMaxRatio}, with \texttt{resourceType} in \{\texttt{PRB\_UL},\texttt{PRB\_DL}\} and members identified by S-NSSAI~\cite{3gpp_ts_28541}.
These attributes realize the standard's Dedicated, Prioritized, and Shared resource behaviors, in which a slice can have a dedicated reservation, preferential use with a guaranteed minimum when needed, or dynamic access to PRBs~\cite{3gpp_ts_28541}.
In TS~28.541~\cite{3gpp_ts_28541}, Prioritized resources are guaranteed to the member when needed and, when idle, may be used by other members of the same managed entity~\cite{3gpp_ts_28541}. This mode of resource sharing, in which resources reserved for one slice may be temporarily used by another slice when idle, is the normative behavior of soft dynamic slicing, not an ad-hoc misconfiguration.

However, the correct intuition treats these policies as incomplete isolation by construction~\cite{elayoubi_5g_ran_slicing_2019,balasingam_application_level_service_2024}. Many implementations are work-conserving~\cite{johnson_nexran_closed_loop_2021}, meaning idle resources may be lent and the operational semantics of a slice floor depend on how the scheduler applies quotas, priorities, and dynamic resource reallocation over time and across traffic directions. Thus, the existence of slice parameters in the management plane leaves instantaneous radio-resource disputes possible for the application~\cite{moreira_noisy_neighbor_influence_2026,akundi_suppressing_noisy_neighbours_2020,vittal_preventing_cross_network_2023}.

\subsection{Asymmetry Between UL and DL}

\begin{figure}[t]
\centering
\includegraphics[width=\columnwidth]{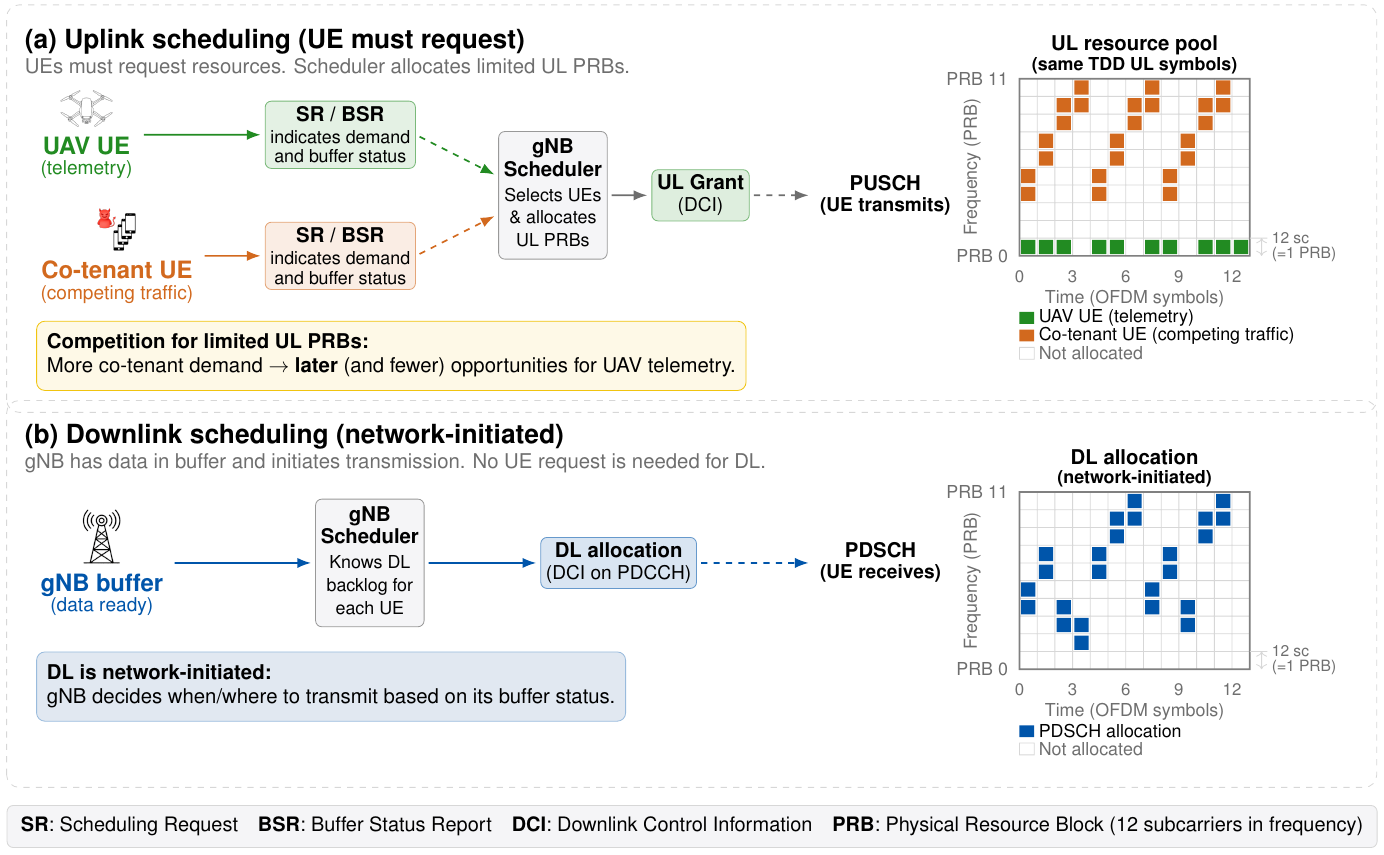}
\caption{UL vs DL scheduling in NR (grant via SR/BSR vs assignment on PDCCH).}
\label{fig:uldlgrants}
\end{figure}

In the 5G New Radio (NR) architecture, UL and downlink (DL) are distinct scheduling problems with different demand origins and backlog visibility. The radio architecture distinguishes demand origins, resource-request mechanisms, and how the scheduler observes backlog and performs allocation~\cite{3gpp_ts_38300}.
On the UL, the UE announces its need through Scheduling Request (SR) and Buffer Status Report (BSR), from which the gNB determines transmission opportunities and allocates grants and PRBs~\cite{3gpp_ts_38321} (Fig.~\ref{fig:uldlgrants}), so co-tenant demand can delay and thin the telemetry opportunities that carry state to the GCS.
On the DL, the gNB transmits scheduling assignments via DCI on the Physical Downlink Control Channel (PDCCH), typically addressed by the UE C-RNTI~\cite{3gpp_ts_38300}.
The relevant asymmetry lies in demand visibility and grant acquisition: the UL is UE-triggered via SR/BSR while the DL relies on gNB PDCCH assignments~\cite{3gpp_ts_38300}, so a healthy DL does not protect freshness on the UL telemetry leg.

Beyond the architectural grant asymmetry, real deployments tend to skew capacity and traffic toward the DL. In existing cellular networks, data traffic is predominantly downlink, whereas UAV application telemetry is sent by uplink~\cite{fotouhi_survey_uav_cellular_2019}, and field-measured commercial plans typically advertise much more downlink than uplink~\cite{baltaci_analyzing_real_time_2022}. Reference TDD configurations in TS~38.101-1~\cite{3gpp_ts_38101}, such as the DDDSU pattern with three DL slots, one special slot, and one UL slot per period, illustrate this temporal asymmetry~\cite{3gpp_ts_38101}, and asymmetric traffic is a relevant characteristic of future mobile systems~\cite{guo_dynamic_tdd_interference_2018}.
The Network Resource Model separates \texttt{PRB\_UL} and \texttt{PRB\_DL} as resource types subject to policy~\cite{3gpp_ts_28541}, which can leave room for incomplete UL enforcement in practice when UL policy is weak or absent.

This difference matters for remote control systems because the telemetry relevant to operator perception originates on the vehicle and rises on the UL~\cite{abdalla_communications_standards_unmanned_2021,allouch_mavsec_securing_mavlink_2019}. DL remains essential for sending commands, but the property at issue here is the freshness of state that returns to the GCS via UL.

\subsection{UAV C2 over 5G}

3GPP has already standardized UAS support and C2 transport over cellular networks, recognizing that BVLOS operations can use the broad coverage, mobility, and network service integration of the 5G architecture~\cite{3gpp_ts_22125,abdalla_communications_standards_unmanned_2021,fotouhi_survey_uav_cellular_2019}. This framing includes cases in which video aids control. Low-latency high-quality video transmissions are essential for safe vehicle operation from the ground~\cite{baltaci_analyzing_real_time_2022}, which brings real UAS closer to a combination of light traffic, such as telemetry and C2 messages, with heavier traffic~\cite{abdalla_communications_standards_unmanned_2021,baguer_enabling_2024}.

However, operationally, the vehicle UE goes beyond a minimal emitter of small state updates~\cite{geraci_what_2022}.
In many plausible profiles, it multiplexes flows with different temporal requirements, making radio scheduler behavior and slice configuration a concrete part of the risk surface~\cite{baguer_enabling_2024}.
This additional flow is control-assistance video feedback, distinct from mission or payload video (for example, surveillance), which TS~22.125 treats separately~\cite{3gpp_ts_22125}.

\subsection{Freshness and AoI}

The real-time systems literature distinguishes average latency from freshness. AoI measures the age of the most recent update delivered to the destination; that is, when the most recent update received at time $t$ has timestamp $u(t)$, the age is $t-u(t)$~\cite{kaul_real_time_status_2012,yates_age_information_introduction_2021}. This formulation is appropriate when losses, queues, and jitter can still allow messages to arrive, but arrive too late to sustain a reliable operational view~\cite{yates_age_information_introduction_2021,tripathi_wiswarm_age_information_2022}. Kaul et al. show that, at the instant of delivery, the age of an update corresponds to its system time, that is, the sum of queue time and service time~\cite{kaul_real_time_status_2012}. Thus, when the UL queue fills under grant deficit, each telemetry sample waits longer, and that wait materializes at the destination as increased delivery age~\cite{kaul_real_time_status_2012,staff_bufferbloat_what_s_2012,jiang_understanding_bufferbloat_cellular_2012}.

For cyber-physical systems, such as UAS, this nuance matters because the destination can receive updates with apparent cadence and still accumulate outdated state~\cite{tripathi_wiswarm_age_information_2022}. For this reason, in BVLOS UAS, freshness is a functional property of remote observation.

\subsection{Motivation: Differential Observability}

Huang et al.\ show that systems can fail partially, ambiguously, or not fully visibly (gray failures), when the observed property diverges from the property that governs the service~\cite{huang_gray_failure_2017}.
Instead of a total, obvious collapse, parts of the system can appear healthy while the function relevant to the user is already degraded. In 5G networks, Xing et al.\ illustrate how known internal mechanisms can generate security consequences when monitoring fixes on the wrong layer~\cite{xing_criticality_2024}.

Applied to UAS over 5G, this reasoning suggests a clear motivation. An operator can observe link latency, availability, and integrity and conclude that connectivity is good. At the same time, the property of real interest, which is state freshness at the GCS, has already failed or is failing~\cite{huang_gray_failure_2017}. The relevance of AoI and proxies of divergence between belief and ground truth arises exactly from this observability gap. Benign experiments on O-RAN 5G for UAS already show that link latency and trajectory error leave residual divergence between what the pilot observes and the received telemetry~\cite{baguer_enabling_2024}, which reinforces why freshness is an important property for monitoring.

\section{Threat Model}
\label{sec:threat}

Given this observability gap, we fix the adversary and the system under test. We consider a BVLOS UAS whose command-and-control (C2) traffic and video feedback traverse the User Plane of a 5G SA network as IP traffic~\cite{3gpp_ts_22125,abdalla_communications_standards_unmanned_2021}.
Although TS~22.125 indicates that C2 communication can include video, we treat C2 messages and video feedback as distinct flows multiplexed on the same UL, aligned with Tables~7.2-1 and~7.2-2 of the standard~\cite{3gpp_ts_22125}.
The GCS and the UAV attach to the same cell and share the S-NSSAI of the critical slice, whereas one or more co-tenant UEs attach to the same cell on a distinct S-NSSAI, a cross-slice design operationally plausible in non-public / campus 5G networks in which several legitimate SIMs share the same coverage and distinct slice policies~\cite{wen_private_2022,pradosgarzon_5g_non_public_2021,5gacia_npn_industrial_2019,vittal_preventing_cross_network_2023}.

The adversary controls only legitimate and authorized UE(s) on the neighbor slice. It does not operate a rogue gNB, perform PHY jamming, compromise MAVLink, the RAN, or the core, and has no administrative privileges beyond what a common authorized subscriber already possesses. We also differ from attacks that exploit the User Plane/GTP-U in the core~\cite{zhang_invade_2025}, since the boundary under test here is UL scheduling contention between S-NSSAIs under soft isolation. FlyBlind does not exploit a flaw in the 5G architecture. The resource sharing it uses is normative~\cite{3gpp_ts_28541}. What breaks is the operational assumption that resource isolation implies state timeliness, and a freshness-sensitive tenant shows it false.

The adversarial capability of co-tenant UEs consists of generating authorized UL traffic at a high average rate, typically in periodic microbursts, to maintain schedulable demand at the gNB. The adversary chooses the number of UEs (dose $N$), the burst pattern, and the start instant after takeoff; the traffic destination is an endpoint external to the victim (for example, a distinct data network pod).

The objective is to degrade state freshness at the GCS, measured by delivery age of position telemetry and by divergence between GCS belief and UAV ground truth. At the same time, monitored connectivity indicators and physical flight remain within apparently healthy ranges. Success is defined as state-estimate divergence in remote situational awareness under an authorized adversary, and not as maximum impact on availability.

We work under soft isolation, without explicit temporal guarantee for the C2 application~\cite{elayoubi_5g_ran_slicing_2019,balasingam_application_level_service_2024}, with the UAV UE multiplexing C2 and video feedback on the same UL link as a realistic operational condition of UAS with visual control assistance~\cite{abdalla_communications_standards_unmanned_2021}. The attack condition is precisely the soft regime, in which idle resources may be shared between slices. In our baseline B0, this corresponds to a UL without \texttt{rRMPolicyRatio} enforcement. Introducing an uplink floor in mitigation configurations constitutes a defense, and not a condition necessary to the attack: from B20 onward, UL \texttt{min\_ratio} restores freshness (Sec.~\ref{sec:mitigations}), reinforcing that the observed effect stems from contention for grants and PRBs. Thus, we explore legitimate resource competition under soft isolation, and not artificial saturation that would produce coarse overload signals.

\begin{figure*}[t]
\centering
\includegraphics[width=\textwidth]{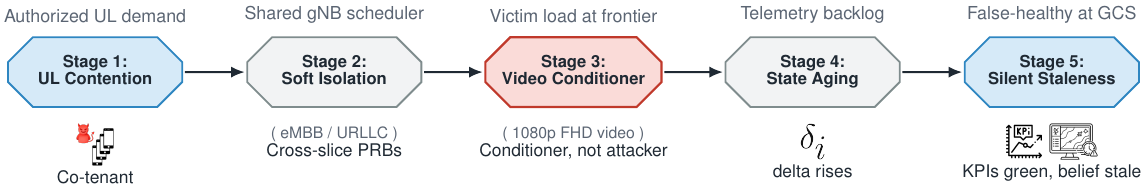}
\caption{FlyBlind pipeline: authorized UL contention $\rightarrow$ Silent State Staleness.}
\label{fig:pipeline}
\end{figure*}

Confidentiality and integrity of C2 content, compromise of network elements and RF attacks, and attacks targeting the victim C2 endpoint are out of scope.
Our axis is temporal, and we do not rely on MAVLink authenticity weaknesses to establish the threat model.

\section{The FlyBlind Attack}
\label{sec:attack}

\subsection{UL Contention Primitive}

FlyBlind does not steal a grant already assigned to the victim. The attacker sustains legitimate schedulable demand on the UL, causing the gNB to split service opportunities among co-tenant UEs and increasing the victim's update wait time. 3GPP specifies, in the MAC (TS~38.321)~\cite{3gpp_ts_38321}, that the BSR procedure provides the serving gNB with information about the volume of UL data in the MAC entity, with logical channels grouped into Logical Channel Groups (LCGs). At the same time, the SR requests UL-SCH resources for new transmissions. In particular, when a Regular BSR is triggered and no UL-SCH resource is available for new transmission, the MAC entity triggers an SR~\cite{3gpp_ts_38321}. Thus, an authorized co-tenant can exploit this SR/BSR$\rightarrow$grant loop without touching the C2 flow. Competition occurs in the shared cell scheduler, and not by injection on the victim IP path.

Figure~\ref{fig:pipeline} compresses that chain: authorized co-tenant demand under soft isolation, with video as load conditioner rather than attacker, ages $\delta_i$ into Silent State Staleness while KPIs stay green and the victim C2 IP remains untouched.
An authorized neighbor competes for grants/PRBs without accessing the victim IP, making UL cross-slice contention operationally relevant in deployments with soft isolation and incomplete UL enforcement (Sec.~\ref{sec:setup}).

\subsection{Video Feedback as Victim-Load Conditioner}

C2 traffic alone consumes little of the UL. Our central observation is that video feedback brings the victim UE close to the temporal boundary at which cross-slice contention begins to age telemetry. We formalize this condition in terms of schedulable demand:
\begin{equation}
\label{eq:load-surface}
L_{\mathrm{attacker}}(N)=N\cdot r_{\mathrm{rogue}},
\qquad
L_{\mathrm{victim}}=r_{\mathrm{C2}}+r_{\mathrm{video}}
\end{equation}
with $r_{\mathrm{rogue}}=20$~Mbps per adversarial UE and $r_{\mathrm{video}}\in\{0,8,12,24\}$~Mbps in the campaign, while $r_{\mathrm{C2}}$ remains light relative to video feedback.
$C_{\mathrm{cell}}$ denotes schedulable UL capacity under the numerology and TDD pattern, and the relation $L_{\mathrm{victim}}+L_{\mathrm{attacker}}$ relative to $C_{\mathrm{cell}}$ characterizes proximity to the service boundary as a capacity relation rather than a direct admitted-PUSCH identity.
Video feedback is not the attack.
It is the load condition under which the attack becomes effective, since, under video ablation ($r_{\mathrm{video}}=0$), multi-rogue contention barely alters telemetry freshness in our experimental regime, whereas, with video at a rate compatible with realistic UAS profiles~\cite{3gpp_ts_22125,baltaci_analyzing_real_time_2022}, the same adversary raises AoI to the order of seconds.

\subsection{False-Healthy State and Silent State Staleness}

We define the security objective in measurable and testable form. In the experimental pipeline, the headline telemetry freshness metric is per-message delivery age. For each \texttt{LOCAL\_POSITION\_NED} correlated between UAV and GCS by the key $(\mathrm{system\_id},\texttt{time\_boot\_ms})$, we measure
\begin{equation}
\label{eq:aoi}
\delta_i
=
t^{\mathrm{mono}}_{\mathrm{GCS},i}
-
t^{\mathrm{mono}}_{\mathrm{UAV},i}
\end{equation}
using \texttt{CLOCK\_MONOTONIC} clocks from the pods~\cite{kaul_real_time_status_2012,yates_age_information_introduction_2021}.
Here, $\texttt{time\_boot\_ms}$ is the message correlation identifier; $\delta_i$ uses a separate monotonic clock.

We report per-message delivery age $\delta_i$, our operational freshness proxy under backlog, aligned with $T_i=t'_i-t_i$ of Kaul et al.\ (system time/age at the instant of delivery) and with the reset $\Delta(t'_j)=t'_j-t_j$ of Yates et al.\ at the destination~\cite{kaul_real_time_status_2012,yates_age_information_introduction_2021}. We report the distributions of $\delta_i$ and \texttt{belief\_vs\_truth} relative to the baseline of the same mission, as operational contrast rather than a universal normative threshold. Under attack, delivery age goes from milliseconds to seconds, and state divergence at the GCS increases by orders of magnitude. At the same time, operational link indicators (GUIDED mode, inactive failsafe, physical tracking) remain close to baseline (Sec.~\ref{sec:eval}). We call Silent State Staleness the phenomenon that FlyBlind induces, a false-healthy state in which telemetry becomes outdated at the GCS while conventional connectivity indicators remain green and the vehicle remains physically intact~\cite{huang_gray_failure_2017}.

We formalize this predicate over the discrete deliveries $i$ of Eq.~\eqref{eq:aoi}, leaving continuous $\Delta(t)$ aside:
\begin{equation}
\label{eq:sss}
\mathrm{SSS}_i
=
H_{\mathrm{net}}
\cdot
H_{\mathrm{veh}}
\cdot
\bigl(1-F_{\mathrm{GCS},i}\bigr)
\end{equation}
where
$F_{\mathrm{GCS},i}=\mathbf{1}\{\delta_i<\tau_{\mathrm{fresh}}\}$
is sample freshness at the GCS,
$H_{\mathrm{net}}=\mathbf{1}\{\mathrm{OWD}_{\mathrm{p99}}<\tau_{\mathrm{owd}}\land\mathrm{avail}>\tau_{\mathrm{avail}}\}$
summarizes conventional health observed on the probe opposite leg (Sec.~\ref{sec:paths}), and
$H_{\mathrm{veh}}=\mathbf{1}\{\texttt{trk\_truth}_{\mathrm{p99}}<\tau_{\mathrm{trk}}\land\mathrm{\texttt{failsafe}}{=}0\land\mathrm{mode}{=}\mathrm{GUIDED}\}$
summarizes physical vehicle health in the evaluation window.
$\mathrm{SSS}_i{=}1$ exactly when network and vehicle appear healthy and the sample at the GCS is stale.
The thresholds $\tau_{\ast}$ are campaign-specific (anchored on baseline and GUIDED/failsafe indicators), and not universal normative thresholds.
Silent State Staleness names the phenomenon in this characterization.
Cadence and freshness can also decouple: packets can continue to arrive while each sample arrives outdated.

This contrast is differential observability in the sense of Huang et al.~\cite{huang_gray_failure_2017}.
A connectivity monitor can report operation while application-state freshness collapses, because link latency/availability/integrity and state age at the destination are distinct properties~\cite{huang_gray_failure_2017}.
Xing et al.\ illustrate, in another 5G domain, how a known mechanism below the application produces a security consequence when monitoring concentrates on the wrong property~\cite{xing_criticality_2024}.
We also report OWD, IP Packet Delay Variation (IPDV), defined as the difference in one-way delay between selected packets~\cite{demichelis_ip_packet_delay_variation_2002}, and availability from a GCS$\rightarrow$UAV probe as the conventional indicator the operator in fact observes on the opposite leg (Sec.~\ref{sec:paths}), and the contrast with freshness on the telemetry leg is precisely what characterizes the false-healthy state.
We use ``silent'' and ``stealth'' in a relative sense with respect to this conventional set, with possible secondary symptoms outside it.
MAVLink heartbeat gaps may appear in runs with $N{=}4$; still, mode remains GUIDED, failsafe remains inactive, and physical tracking remains at baseline.

\begin{figure}[t]
\centering
\includegraphics[width=0.90\columnwidth]{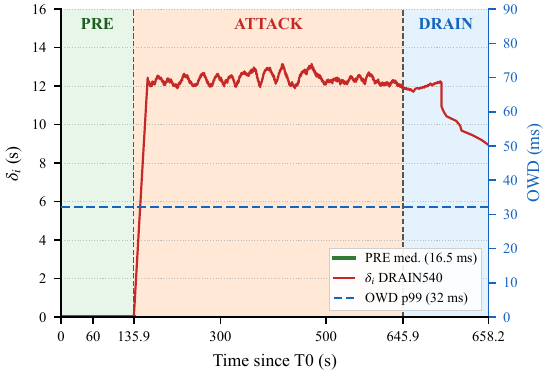}
\caption{PRE/ATTACK/DRAIN temporal signature ($\delta_i$ fill/saturate/drain). DRAIN540 identifies the attack run with a 540~s trajectory.}
\label{fig:pre-attack-drain}
\end{figure}

The underlying queueing mechanics have been known since the classic bufferbloat discussion, where excess buffering produces high latency without necessarily bringing down the path through massive loss~\cite{staff_bufferbloat_what_s_2012,jiang_understanding_bufferbloat_cellular_2012}.
The contribution is to show the security consequence of Silent State Staleness under cross-slice soft isolation, with the vehicle intact and operational connectivity preserved.
The causal chain is that of a queue operating under a service deficit, where grant contention reduces UL opportunities for the victim, UL backlog grows, $\delta_i$ rises and saturates, and drains when the attack ceases (Fig.~\ref{fig:pre-attack-drain})~\cite{kaul_real_time_status_2012,staff_bufferbloat_what_s_2012}.
The PRE/ATTACK/DRAIN signature, with OWD p99 stuck in the tens of milliseconds while $\delta_i$ fills and drains with attack onset and termination, rules out a constant clock offset: the median offset between pod monotonic clocks stays on the order of sub-millisecond to a few milliseconds, which makes this temporal signature incompatible with a static measurement bug.

\subsection{LCG, Internal Head-of-Line (HOL), and What the Claim Admits}
\label{sec:lcg-hol}

BSR reports UL buffer status per LCG, aggregating logical channels~\cite{3gpp_ts_38321}.
The causal chain we observe begins in authorized UL cross-slice contention, which reduces the UL service rate granted to the victim UE and ages telemetry in a multiplexed C2+video feedback queue on the same data radio bearer (DRB). In our stack, C2 and video feedback share the same UE and the same PDU Session on slice B under 5QI~82 provisioning. Thus, internal HOL at the UE constitutes the final stage of this chain. Grant contention is the origin. Video feedback is the load conditioner, and not the attacker. It places the UE at the operational boundary at which contention begins to produce temporal impact. Ablation under video-off shows that a high NPRB share (the OAI/ORANSlice PRB allocation share) alone leaves false-healthy absent in our regime (Tab.~\ref{tab:video-feedback-ablation}, $N{=}4$, vid.\ 0, share A $95.3\%$, $\delta$ p99 $65.3$~ms).

Outdated information originates from UL cross-slice grant contention, with internal HOL at the UE as the final stage of this chain. This attribution is supported by evaluation (Sec.~\ref{sec:eval}), in which video feedback ablation, the \texttt{max\_sched\_ues} sweep, width sensitivity, and the reversible PRE/ATTACK/DRAIN signature converge on the same origin. These results rule out a static internal queue, the scheduling cap, and a static clock offset as dominant causes.

The security property is independent of this final stage. A co-tenant without administrative privileges and without touching the C2 flow produces a false-healthy invisible to monitored KPIs, with measurable cyber-physical harm at a realistic UAS operating point with video on the UL~\cite{baltaci_analyzing_real_time_2022}.
This distinguishes FlyBlind from classic bufferbloat, which does not formalize cross-slice induction by an authorized adversary nor the false-healthy predicate with GCS-vs-truth evidence~\cite{jiang_understanding_bufferbloat_cellular_2012,staff_bufferbloat_what_s_2012}, and the resource sharing FlyBlind exploits is the normative behavior of soft dynamic slicing rather than an ad-hoc misconfiguration.
RLC/PDCP/LCG buffers remain present in this final stage~\cite{3gpp_ts_38300,3gpp_ts_38321}, but cross-slice-induced grant deficit is the origin that feeds this backlog.

\subsection{Execution}

Execution is simple. Rogue UEs attach, establish a PDU Session on slice A, and generate periodic UL microbursts to an external destination, while dose $N$ controls how many UEs compete.
The attack starts after takeoff (to separate arm/takeoff from contention), and the vehicle executes a GUIDED circular orbit to remain under the same cell and isolate handover and coverage variation, so the causal variable under test remains UL scheduling contention.

\section{Experimental Setup}
\label{sec:setup}

\subsection{Stack and Slices}

The testbed follows a standard and fully reproducible 5G SA architecture, running under Kubernetes orchestration~\cite{kubernetes} (Kind~\cite{kind}).
For Core and RAN we use the OpenAirInterface solution~\cite{openairinterface}, an open-source stack widely used in 5G experimentation.
RAN slicing uses ORANSlice, which extends the OAI proportional-fair scheduler with a slice-aware two-tier scheme aligned with \texttt{rRMPolicyMinRatio}/\texttt{rRMPolicyMaxRatio} per S-NSSAI~\cite{cheng_oranslice_open_source_2024}.

The radio interface is OAI rfsim~\cite{oai_rfsimulator}, band n78 (TDD), 106 and 51~PRB, numerology $\mu=1$, using the same class of n78 configurations adopted in several published O-RAN testbeds~\cite{cheng_oranslice_open_source_2024,villa_x5g_open_programmable_2025,3gpp_ts_38101}.
In our configuration, rfsim emulates the radio interface deterministically, which isolates UL scheduling as the variable under test.
External RF validity is discussed in threats to validity (Sec.~\ref{sec:discussion}).

For UAS, we use ArduPilot~\cite{ardupilot} SITL (arducopter) with MAVLink/UDP on the User Plane~\cite{mavlink_guide,pymavlink,ardupilot_sitl,ardupilot_simulation}, closing the cyber-physical loop of modes, failsafes, and physical tracking~\cite{baldi_ardupilot_based_adaptive_2022,konapalli_reverse_engineering_control_2025,allouch_mavsec_securing_mavlink_2019}.

The two S-NSSAIs share the cell, gNB, and, in this deployment, the same UPF.
Slice A (rogues) uses SST$=$1, SD$=$\texttt{0xFFFFFF}, DNN \texttt{oai}, and 5QI~9, while slice B (GCS+UAV) uses SST$=$1, SD$=$\texttt{0x000001}, DNN \texttt{oai.urllc}, and 5QI~82.
All experimental evidence in this paper is cross-slice, with a separate mitigation configuration.

In the experimental stack, UL is cell-wide PF under soft isolation (\texttt{min\_ratio} A$=$0 and B$=$0), and UL protection by MinRatio on slice B enters only in the mitigation fork (Sec.~\ref{sec:mitigations}), anchoring the class of deployments with incomplete or asymmetric UL enforcement~\cite{3gpp_ts_28541,elayoubi_5g_ran_slicing_2019}.

\subsection{Instrumentation and Metrics}
\label{sec:paths}

\begin{figure}[t]
\centering
\includegraphics[width=\columnwidth]{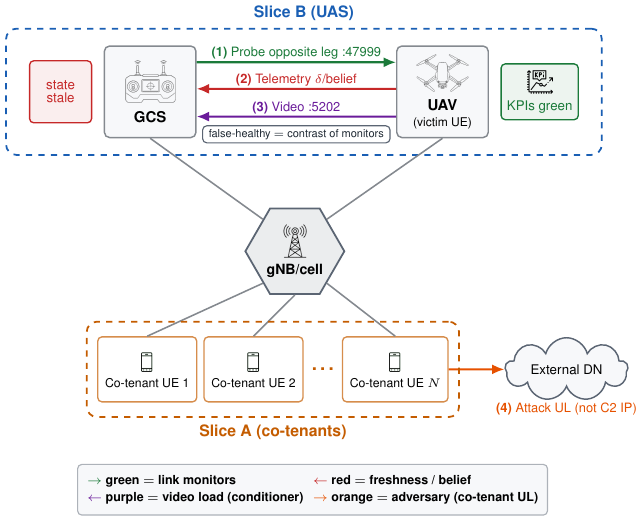}
\caption{Instrumented paths (probe vs telemetry).}
\label{fig:paths}
\end{figure}

Figure~\ref{fig:paths} separates the paths that false-healthy contrasts: the GCS$\rightarrow$UAV probe on :47999 that the operator monitors, the UAV$\rightarrow$GCS telemetry/belief leg that governs situational awareness, video feedback on :5202, and attack UL to an external DN.
IP availability is not message cadence, and cadence is not state freshness~\cite{yates_age_information_introduction_2021,huang_gray_failure_2017}.
The OWD/IPDV/availability probe matches the conventional monitor on the opposite leg from UL telemetry, so green path KPIs can coexist with a stale GCS belief.
Telemetry \texttt{LOCAL\_POSITION\_NED} and delivery age $\delta_i$ (Eq.~\eqref{eq:aoi}) run on the UAV$\rightarrow$GCS leg.

Delivery timestamps use local \texttt{CLOCK\_MONOTONIC} from the UAV and GCS pods.
\texttt{time\_boot\_ms} correlates the same \texttt{LOCAL\_POSITION\_NED} between UAV egress and GCS arrival, and serves only as a correlator for $\delta_i$.
The median offset between monotonic clocks, measured by handshake, typically stays below $1$~ms, while $\delta_i$ under attack enters the scale of seconds.

We report (i)~delivery age $\delta_i$ of telemetry~\cite{kaul_real_time_status_2012,yates_age_information_introduction_2021,tripathi_wiswarm_age_information_2022}; (ii)~\texttt{belief\_vs\_truth} as an operational proxy of state divergence; (iii)~\texttt{trk\_err}/\texttt{trk\_truth}/\texttt{xtrack}; (iv)~OWD, IPDV, and availability from the probe; (v)~heartbeat gaps, flight mode, and failsafe; (vi)~NPRB share per slice and observed absolute UL occupancy; and (vii)~offered/sender from the rogues and absolute UL from the UAV.
We aggregate quantiles per run and report median and spread across runs, treating 10~Hz telemetry samples as dependent within a run rather than as independent and identically distributed (i.i.d.)\ in a single pool.

\label{sec:trk-err}
The testbed report uses \texttt{trk\_err} for the error between the believed position at the GCS and the trajectory plane, distinct from the physical deviation of the vehicle.
Let $\mathbf{p}_{\mathrm{plan}}(t)$ be the reference point of the GUIDED orbit and $\mathbf{p}_{\mathrm{GCS}}(t)$ the latest \texttt{LOCAL\_POSITION\_NED} at the GCS.
We define
\begin{equation}
\label{eq:trk-err}
\mathrm{trk\_err}(t)
=
\bigl\|
\mathbf{p}_{\mathrm{GCS}}(t)
-
\mathbf{p}_{\mathrm{plan}}(t)
\bigr\|_{2}.
\end{equation}
In contrast, \texttt{trk\_truth} uses the SITL egress position $\mathbf{p}_{\mathrm{truth}}(t)$:
\begin{equation}
\label{eq:trk-truth}
\mathrm{trk\_truth}(t)
=
\bigl\|
\mathbf{p}_{\mathrm{truth}}(t)
-
\mathbf{p}_{\mathrm{plan}}(t)
\bigr\|_{2}.
\end{equation}
These metrics, GUIDED mode, and failsafe derive from ArduPilot firmware in SITL~\cite{ardupilot_sitl,ardupilot_simulation}.
The \texttt{belief\_vs\_truth} proxy, reconstructed offline from \texttt{.csv} files, measures state divergence at the granularity of instrumented arrivals.
For the circular orbit, cross-track error (xtrack) anchors physical tracking.

\subsection{PDU Session and QoS Provisioning}
\label{sec:bearers}

GCS and UAV use slice B (DNN \texttt{oai.urllc}, 5QI~82), and the rogues use slice A (DNN \texttt{oai}, 5QI~9).
C2 and video feedback are carried through the same UAV UE and the same PDU Session under 5QI~82 (Delay-critical GBR)~\cite{3gpp_ts_23501}, with Session-AMBR UL configured at 5~Mbps.
In this OAI/ORANSlice deployment, Session-AMBR functions as a provisioning label and the UL is not policed to that ceiling, which instantiates the incomplete UL enforcement class of the threat model (Sec.~\ref{sec:threat}).
Deployments that enforce GBR/AMBR policing would narrow the attack window and delimit the class under study~\cite{3gpp_ts_23501}.
The LCG/internal HOL chain of this multiplexing is in Sec.~\ref{sec:lcg-hol}.

\subsection{Conventional KPIs Versus Freshness}

Industry monitors the C2 link by latency, availability, and integrity. Honeywell TestReport-263, aligned with DO-377A, reports latency of 1.0~s at least 95\% of the time and serves as a concrete measurement reference, without claiming normative compliance~\cite{honeywelltestreport263do377c2linktest_us_2023}.
The MIT-LL AD1197090 report anchors the same KPI family in the DO-377A ecosystem~\cite{mitllad1197090c2linksrmanalysisdo377a_us_2022}.
In 5G, TS~28.554 fixes end-to-end KPIs in terms of latency, reliability, and related service metrics~\cite{3gpp_ts_28554}, while TS~22.125 defines UAS C2 and video KPIs in terms of latency, reliability, and data rate~\cite{3gpp_ts_22125}, and Yates et al.\ and WiSwarm position AoI as a timeliness metric at the destination~\cite{yates_age_information_introduction_2021,tripathi_wiswarm_age_information_2022}.

To make the operational contrast, we define freshness-aware availability over the $n$ correlated telemetry deliveries:
\begin{equation}
\label{eq:afresh}
A_{\mathrm{fresh}}
=
\frac{1}{n}
\sum_{i=1}^{n}
\mathbf{1}\{\delta_i < \tau_{\mathrm{fresh}}\}
\end{equation}
with the same $\delta_i$ from Eq.~\eqref{eq:aoi}.
We fix $\tau_{\mathrm{fresh}}{=}150$~ms as the operational threshold of the campaign, above baseline $\delta$ p99 (${\approx}56$~ms) and well below the attack regime in seconds, without claiming a DO-377 normative threshold.
When $H_{\mathrm{net}}{=}H_{\mathrm{veh}}{=}1$ in the window, Eq.~\eqref{eq:sss} reduces to $\mathrm{SSS}_i{=}1-F_{\mathrm{GCS},i}$, and the mass of Silent State Staleness coincides with $1-A_{\mathrm{fresh}}$.

\subsection{Load Model}
\label{sec:rogue-load}

Video load is video-rate-matched load: a \texttt{iperf3}~\cite{iperf3} UDP flow UAV$\rightarrow$GCS at rates compatible with published bitrates for video used to aid UAV control~\cite{3gpp_ts_22125,baltaci_analyzing_real_time_2022,abdalla_communications_standards_unmanned_2021}.
The constant-rate, rate-matched load isolates the temporal boundary under controlled offered demand.
12~Mbps sits within the range reported for real-time video in remote piloting~\cite{baltaci_analyzing_real_time_2022}.
Rogues generate authorized UL microbursts (SR/BSR$\rightarrow$grant) to an external destination.
With $r_{\mathrm{rogue}}=20$~Mbps, Eq.~\eqref{eq:load-surface} gives $L_{\mathrm{attacker}}(1)=20$~Mbps and $L_{\mathrm{attacker}}(4)=80$~Mbps.
Table~\ref{tab:rogue-load} shows NPRB share A rising and UAV UL falling under attack while Offered diverges from admitted load, so scheduler contention is readable from these proxies without requiring PUSCH saturation (106~PRB, 600~s, three runs).

\begin{table}[t]
\centering
\caption{UL load and scheduling proxies.}
\label{tab:rogue-load}
\scriptsize
\setlength{\tabcolsep}{0.4pt}
\renewcommand{\arraystretch}{0.95}
\begin{tabular*}{\columnwidth}{@{\extracolsep{\fill}}lrrrrrr@{}}
\toprule
Condition & Off. & Snd. & UAV & Cell & Sh.A & $\delta$ p99 \\
 & (Mbps) & (Mbps) & (Mbps) & (Mbps) & (\%) &  \\
\midrule
Baseline, vid.\ 12 & -- & -- & $12.50{\pm}0.00$ & $13.28{\pm}0.01$ & $24.70{\pm}0.47$ & $56.44{\pm}2.97$~ms \\
Atk $N{=}1$, vid.\ 12 & 20 & 19.9 & $12.51{\pm}0.00$ & $13.40{\pm}0.00$ & $59.90{\pm}1.60$ & $58.53{\pm}2.59$~ms \\
Atk $N{=}4$, vid.\ 0 & 80 & 79.5 & $0.72{\pm}0.01$ & $0.89{\pm}0.01$ & $95.26{\pm}0.11$ & $65.35{\pm}2.17$~ms \\
Atk $N{=}4$, vid.\ 12 & 80 & 79.5 & $8.04{\pm}0.06$ & $8.82{\pm}0.07$ & $76.56{\pm}0.37$ & $12.16{\pm}0.17$~s \\
\bottomrule
\end{tabular*}
\end{table}

We report offered, sender, Cell UL, and share jointly (Offered $\neq$ sender $\neq$ admitted $\neq$ PUSCH).
NPRB share indexes grant opportunities in the scheduler, as distinct from MAC bytes or PUSCH saturation.
In the baseline, share A of $24.7\%$ with idle rogues illustrates this intentional difference between scheduler presence and useful throughput.

\subsection{Exposed Parameters and Aggregation}

The OAI parameter \texttt{max\_sched\_ues} defines the default of ${\approx}4$ UEs per scheduling opportunity at 106~PRB.
Headline magnitude under variation of this cap is reported in Evaluation.

\section{Evaluation}
\label{sec:eval}

Unless otherwise indicated, we report the median across three valid runs (600~s, 106~PRB, cross-slice or CROSS topology).
Attack onset occurs after takeoff, during the GUIDED orbit on the same cell.

\subsection{False-Healthy Headline}

Table~\ref{tab:false-healthy} establishes the falsifiable predicate under attack $N{=}4$ with video feedback at 12~Mbps. $H_{\mathrm{net}}$ and $H_{\mathrm{veh}}$ stay green while $F_{\mathrm{GCS}}$ fails on $\delta$ and \texttt{belief\_vs\_truth}, so operational safety is not link health.
Divergence between belief and ground truth supplies the pilot side reading that audits situational awareness under authorized co-tenancy.

\begin{table}[tb]
\centering
\caption{False-healthy predicate under FlyBlind.}
\label{tab:false-healthy}
\scriptsize
\setlength{\tabcolsep}{4pt}
\renewcommand{\arraystretch}{0.95}
\begin{tabular*}{\columnwidth}{@{\extracolsep{\fill}}lrr@{}}
\toprule
Metric & Baseline & Atk $N{=}4$ \\
\midrule
\multicolumn{3}{@{}l@{}}{Network (probe opposite leg)} \\
OWD p99 (ms) & $27.77{\pm}0.22$ & $29.66{\pm}0.22$ \\
IPDV p99 (ms) & $8.95{\pm}0.07$ & $9.71{\pm}0.08$ \\
Availability (\%) & $100.00{\pm}0.00$ & $99.96{\pm}0.04$ \\
\midrule
\multicolumn{3}{@{}l@{}}{Vehicle (physical tracking)} \\
\texttt{trk\_truth} p99 (m) & $9.28{\pm}0.01$ & $9.29{\pm}0.03$ \\
xtrack steady p99 (m) & $0.29$ & $0.29$ \\
Flight mode & GUIDED & GUIDED \\
Failsafe & none & none \\
\midrule
\multicolumn{3}{@{}l@{}}{Freshness / belief at the GCS} \\
Delivery age $\delta$ p50 & $16.62{\pm}0.38$~ms & $11.62{\pm}0.13$~s \\
Delivery age $\delta$ p99 & $56.44{\pm}2.97$~ms & $12.16{\pm}0.17$~s \\
\texttt{belief\_vs\_truth} p99 (m) & $1.07{\pm}0.01$ & $45.67{\pm}0.67$ \\
\texttt{trk\_err} p99 (m) & $9.35{\pm}0.01$ & $50.86{\pm}0.33$ \\
Heartbeat gaps ($>$1.5~s) & 0 & 32 \\
Max heartbeat gap (s) & --- & $\approx$2.9--3.2 \\
\bottomrule
\end{tabular*}
\end{table}

Under attack, the GCS--UAV link remains operational, with GUIDED mode preserved, inactive failsafe, and \texttt{trk\_truth}/\texttt{xtrack} close to baseline.
Meanwhile, delivery age $\delta$ jumps from tens of milliseconds to $11.62$/$12.16$~s (p50/p99), and \texttt{belief\_vs\_truth} p99 goes from $1.07$~m to $45.67$~m.
The probe on the opposite leg remains close to baseline, with OWD p99 of $29.66$~ms and availability of $99.96\%$, serving only as contrast with conventional monitors (Sec.~\ref{sec:paths}).
With $\tau_{\mathrm{fresh}}{=}150$~ms in Eq.~\eqref{eq:afresh}, $A_{\mathrm{fresh}}$ goes from $100\%$ at baseline to ${\approx}0.69\%$ under attack (median across runs, from \texttt{miss\_ratio\_delay} in summaries), while $A_{\mathrm{avail}}{\approx}99.96\%$.
Under $H_{\mathrm{net}}{=}H_{\mathrm{veh}}{=}1$ in that window, Eq.~\eqref{eq:sss} classifies most deliveries as $\mathrm{SSS}_i{=}1$.
Heartbeat gaps rise from 0 to 32 (median), with max gap on the order of $2.9$ to $3.2$~s.
Thus, secondary symptoms are present, but they leave mode GUIDED and failsafe inactive, while state divergence at the GCS remains large.

\subsection{Video Feedback as Conditioner}

Table~\ref{tab:video-feedback-ablation} fixes $N{=}4$ and varies the video-rate-matched load.
With vid.\ 0, $\delta$ stays in milliseconds despite high share A, so video is the frontier conditioner and the co-tenant remains the attacker.

\begin{table}[t]
\centering
\caption{Video feedback ablation under attack.}
\label{tab:video-feedback-ablation}
\scriptsize
\setlength{\tabcolsep}{3pt}
\begin{tabular*}{\columnwidth}{@{\extracolsep{\fill}}lrrrr@{}}
\toprule
Condition & $\delta$ p99 & belief p99 & UAV UL & Share A \\
 &  & (m) & (Mbps) & (\%) \\
\midrule
Atk $N{=}4$, vid.\ 0 & $65.35{\pm}2.17$~ms & $1.07{\pm}0.00$ & $0.72{\pm}0.01$ & $95.26{\pm}0.11$ \\
Atk $N{=}4$, vid.\ 8 & $12.24{\pm}0.16$~s & $45.67{\pm}0.62$ & $7.24{\pm}0.01$ & $76.87{\pm}0.08$ \\
Atk $N{=}4$, vid.\ 12 & $12.16{\pm}0.17$~s & $45.67{\pm}0.67$ & $8.04{\pm}0.06$ & $76.56{\pm}0.37$ \\
Atk $N{=}4$, vid.\ 24 & $12.44{\pm}0.26$~s & $46.94{\pm}0.62$ & $10.28{\pm}0.06$ & $75.90{\pm}0.13$ \\
\bottomrule
\end{tabular*}
\end{table}

With vid.\ 0, $\delta$ p99 remains at $65.3$~ms and belief p99 at $1.07$~m, despite share A of $95.3\%$.
With 8/12/24~Mbps, $\delta$ remains on the order of $12$~s and belief p99 between approximately $45$ and $47$~m.
After crossing this threshold, increasing bitrate barely alters the regime, which suggests service deficit and queue instability rather than linear degradation with throughput.

The cause under test is the UL scheduler of the gNB; downstream User Plane elements remain secondary in this attribution.
Although the two slices share the same UPF in this deployment, three results point to the RAN as the dominant cause, with UPF and host saturation ruled out by the controls below.

First, UL MinRatio on slice B restores freshness under the same UPF and the same User Plane path (Sec.~\ref{sec:mitigations}). Second, under video ablation, the same adversarial dose barely alters $\delta_i$, despite the high NPRB share of slice A. Third, under attack with $N{=}4$, median max UPF CPU is $5.20\%$ (spread $0.60$) and median max host CPU busy is $42.38\%$ (spread $1.63$), with median max \texttt{loadavg1} at $9.80$ (spread $1.19$), against $6.40\%$ / $39.42\%$ / $8.55$ at baseline (\texttt{nproc}${=}16$, validation threshold $14$).

Thus, freshness degradation occurs while UPF CPU and host queue remain below saturation.
The causal attribution is PRB/grant contention in the UL scheduler.
Path congestion would saturate the shared User Plane and respond to controls at the UPF or host, while an uplink floor at the RAN restores freshness under the same UPF and the same path.
Under attack, UAV UL falls to $8.04$~Mbps and remains positive.
Meanwhile, $\delta_i$ saturates and drains, while the opposite-leg probe remains close to baseline.
This signature is consistent with grant opportunity deficit under a still-live IP path.

\begin{figure}[H]
\centering
\includegraphics[width=0.75\columnwidth]{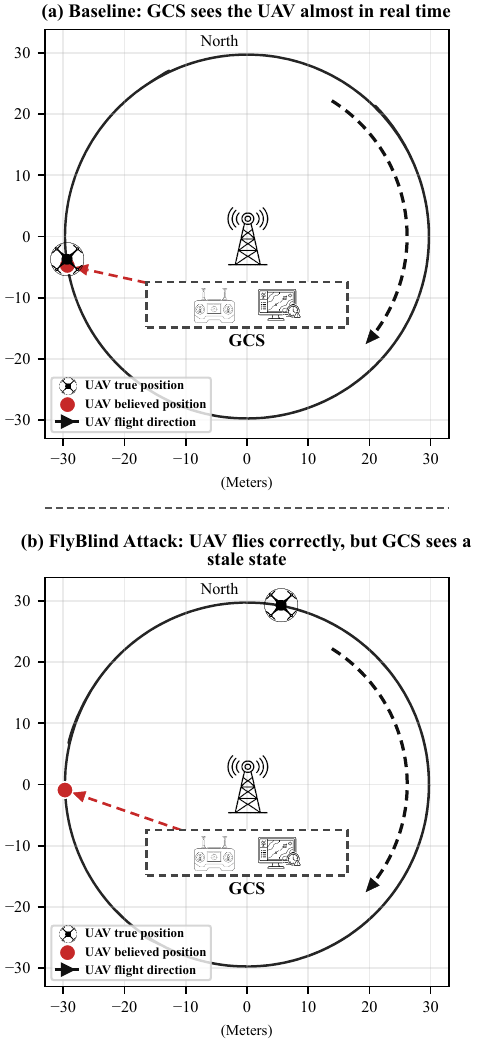}
\caption{Circular UAV trajectory: truth vs belief at the GCS.}
\label{fig:orbit}
\end{figure}

\subsection{Cyber-Physical Divergence}

The geometric block of Table~\ref{tab:false-healthy} isolates the cyber-physical trio. Apparent tracking error at the GCS grows while physical deviation stays near baseline, so the cost is operational blindness with intact flight.
Figure~\ref{fig:orbit} shows the cyber-physical cost on the GUIDED orbit: under attack the vehicle stays on trajectory while GCS belief freezes into a ghost position, so \texttt{trk\_err} and belief-vs-truth rise with physical flight intact~\cite{yates_age_information_introduction_2021,huang_gray_failure_2017,baldi_ardupilot_based_adaptive_2022}.

Under attack, \texttt{trk\_err} and \texttt{belief\_vs\_truth} increase jointly, while \texttt{trk\_truth} and \texttt{xtrack} remain close to baseline.
The jump in \texttt{trk\_err} is induced by freshness loss, when UL contention delays updates, $\mathbf{p}_{\mathrm{GCS}}$ freezes or updates with delayed samples while $\mathbf{p}_{\mathrm{plan}}(t)$ continues to advance on the orbit, so that~\eqref{eq:trk-err} grows with the vehicle physically on the trajectory.
Arrival gaps and GUIDED orbit speed amplify the same effect, and the adversarial dose conditions $\delta_i$ and, therefore, \texttt{trk\_err}.
The difference $\mathrm{p99}(\mathrm{trk\_err})-\mathrm{p99}(\mathrm{trk\_truth})$ summarizes this apparent jump at the GCS and complements the \texttt{belief\_vs\_truth} proxy as a parallel observable. We treat this condition as enabling an induced error condition in remote situational awareness, whose operational impact depends on mission window and operator policy. For example, at the measured belief divergence of about 46~m, an operator issuing a deconfliction or geofence decision from the GCS map acts on a position the aircraft has already left, so a maneuver that reads as clear on the displayed state can bring the vehicle within the separation margin of an obstacle or of nearby traffic.

\subsection{Dose Sweep Under Video Feedback}

Table~\ref{tab:dose-sweep} varies $N$ with video feedback fixed at 12~Mbps.
Dose $N{=}1$ stays near baseline freshness while $N{\ge}2$ produces a monotonic increase in staleness, so the adversarial effect depends on dose, with the near-null effect at $N{=}1$ anchoring the low-dose regime.

\begin{table}[t]
\centering
\caption{Dose sweep under video feedback.}
\label{tab:dose-sweep}
\scriptsize
\setlength{\tabcolsep}{2pt}
\renewcommand{\arraystretch}{0.95}
\begin{tabular*}{\columnwidth}{@{\extracolsep{\fill}}lrrrrr@{}}
\toprule
$N$ & Offered & $\delta$ p99 & belief p99 & \texttt{trk\_err} p99 & Share A \\
 & (Mbps) &  & (m) & (m) & (\%) \\
\midrule
1 & 20 & $58.53{\pm}2.59$~ms & $1.07{\pm}0.00$ & $9.35{\pm}0.01$ & $59.90{\pm}1.60$ \\
2 & 40 & $7.48{\pm}0.16$~s & $30.38{\pm}0.92$ & $37.76{\pm}0.71$ & $63.41{\pm}0.20$ \\
3 & 60 & $9.95{\pm}0.29$~s & $38.98{\pm}0.80$ & $45.24{\pm}0.73$ & $71.37{\pm}0.02$ \\
4 & 80 & $12.16{\pm}0.17$~s & $45.67{\pm}0.67$ & $50.86{\pm}0.33$ & $76.56{\pm}0.37$ \\
\bottomrule
\end{tabular*}
\end{table}

At $N{=}1$, $\delta$ p99 ($58.5$~ms) and belief ($1.07$~m) remain on the order of baseline, despite share A of $59.9\%$.
The transition $N{=}1\rightarrow N{=}2$ is abrupt. $\delta$ p99 jumps from tens of milliseconds to $7.48$~s, indicating a regime change under the same video-rate-matched load.
From $N{=}2$ onward, belief and \texttt{trk\_err} increase monotonically up to $N{=}4$.
This transition coincides with approaching the \texttt{max\_sched\_ues} limit ${\approx}4$ at 106~PRB when the number of active rogues goes from one to two.
We separate this effect by the near-null at $N{=}1$, by the PRE/ATTACK/DRAIN signature (Tab.~\ref{tab:pre-attack-drain}), and by sensitivity at 51~PRB (Tab.~\ref{tab:prb-regime}); the \texttt{max\_sched\_ues} cap sweep fixes $N{=}4$ and serves a different control.

In the OAI/ORANSlice scheduler used on the testbed, the internal \texttt{max\_sched\_ues} limit derives from $\lfloor\mathrm{BW}/(4\cdot 6)\rfloor$, truncated by \texttt{MAX\_DCI\_CORESET}, which results in $4$ UEs per round at 106~PRB and $2$ at 51~PRB.
Cross-slice experiments keep 6 UEs attached (4 rogues + GCS + UAV), and $N$ controls how many rogues attack actively.
With six UEs attached, raising \texttt{max\_sched\_ues} from 4 to 8 allows all to be considered in the slot under cap-free rotation.
Still, $\delta$ p99 remains at $12.62$/$12.69$/$12.55$~s for caps 4/6/8 (1 rep per cap), so headline magnitude tracks UL contention across the measured caps.
Video feedback ablation ($N{=}4$, vid.\ 0) and restoration by UL MinRatio show that incomplete soft isolation and victim load continue to condition the effect under the same scheduler implementation.

\subsection{Bandwidth Sensitivity (51 vs 106 PRB)}
\label{sec:eval-prb-regime}

Table~\ref{tab:prb-regime} contrasts the same $N{=}4$ attack with 8~Mbps video feedback under 106- and 51-PRB configurations.
The narrower regime aggravates $\delta$ and belief ($\delta$ p99 $15.81$~s and belief $54.14$~m at 51~PRB) while OWD and tracking stay green, so the phenomenon persists and worsens under fewer UL symbols rather than depending on one bandwidth.
False-healthy appears at both 51 and 106~PRB, consistent with fewer grant opportunities and \texttt{max\_sched\_ues}${\approx}2$ at 51~PRB.

\begin{table}[t]
\centering
\caption{106 vs 51~PRB regime sensitivity.}
\label{tab:prb-regime}
\scriptsize
\setlength{\tabcolsep}{4pt}
\begin{tabular*}{\columnwidth}{@{\extracolsep{\fill}}lrr@{}}
\toprule
Metric & 106~PRB & 51~PRB \\
\midrule
OWD p99 (ms) & $29.78{\pm}0.13$ & $21.21{\pm}0.13$ \\
Availability (\%) & $99.97{\pm}0.07$ & $100.00{\pm}0.00$ \\
Delivery age $\delta$ p99 & $12.24{\pm}0.16$~s & $15.81{\pm}0.34$~s \\
\texttt{belief\_vs\_truth} p99 (m) & $45.67{\pm}0.62$ & $54.14{\pm}0.81$ \\
UAV UL (Mbps) & $7.24{\pm}0.01$ & $5.93{\pm}0.04$ \\
Share A (\%) & $76.87{\pm}0.08$ & $75.60{\pm}0.15$ \\
xtrack steady p99 (m) & $0.29$ & $0.29$ \\
Failsafe & none & none \\
\bottomrule
\end{tabular*}
\end{table}

\subsection{Temporal Signature (PRE/ATTACK/DRAIN)}
\label{sec:temporal-signature}

Table~\ref{tab:pre-attack-drain} closes temporal causality for the headline ($N{=}4$, vid.\ 12, 106~PRB).
PRE stays in milliseconds, ATTACK saturates near $12$~s, DRAIN decays gradually ($\delta$ p50 $10.81$~s vs.\ $12.31$~s in ATTACK), and OWD p99 stays near $32$~ms, so the signature is queue drain under grant deficit while the link monitor misses the freshness collapse.
PRE comes from baseline vid.\ 12 (median of three runs). ATTACK and DRAIN come from run DRAIN540 (trajectory 540~s, capture until after \texttt{attack\_end}).
The PRE/ATTACK/DRAIN shape reinforces attribution in Sec.~\ref{sec:lcg-hol}, with cross-slice grant contention and internal HOL as the final stage of this chain.

\begin{table}[b]
\centering
\caption{Freshness by PRE, ATTACK, and DRAIN phase.}
\label{tab:pre-attack-drain}
\scriptsize
\setlength{\tabcolsep}{3pt}
\begin{tabular*}{\columnwidth}{@{\extracolsep{\fill}}lrrr@{}}
\toprule
Metric & PRE & ATTACK & DRAIN \\
\midrule
$\delta$ p50 & $16.53$~ms & $12.31$~s & $10.81$~s \\
$\delta$ p99 & $19.81$~ms & $13.02$~s & $12.22$~s \\
belief p99 (m) & $0.63$ & $48.27$ & $45.62$ \\
OWD p99 (ms) & $27.77$ & $32.17$ & $32.17$ \\
Tel.\ gap p99 (ms) & $254.9$ & $521.7$ & $341.4$ \\
Heartbeat gaps ($>$1.5~s) & $0$ & $35$ & n/a \\
GUIDED / failsafe & GUIDED / none & GUIDED / none & GUIDED / none \\
\bottomrule
\end{tabular*}
\end{table}

\section{Mitigations and the Cost of Isolation}
\label{sec:mitigations}

We ask which mitigations alter this behavior and at what cost. We frame these mitigations as causal identification.
Each one answers a hypothesis about the origin of aging, whether local host queue, co-tenant dose, or absence of uplink floor on the victim slice.

\subsection{Isolation Cost Metrics}
\label{sec:reservation-cost}

When the mitigation fork activates UL protection on slice B, we measure freshness restoration and the cost of isolation.
Two metrics anchor the classic tradeoff in which strong isolation reduces statistical efficiency.

Reservation Cost quantifies the relative loss of cell throughput when reservation is active:
\begin{equation}
\label{eq:reservation-cost}
\mathrm{ReservationCost}
=
\frac
{T_{\mathrm{cell}}^{\mathrm{(no)}}-T_{\mathrm{cell}}^{\mathrm{(rsv)}}}
{T_{\mathrm{cell}}^{\mathrm{(no)}}}
\end{equation}
where $T_{\mathrm{cell}}$ is aggregate cell throughput.

Reservation Waste quantifies the fraction of reserved UL PRBs left unused by the victim:
\begin{equation}
\label{eq:reservation-waste}
\mathrm{ReservationWaste}
=
\frac{U_{\mathrm{rsv}}}{R_{\mathrm{UL}}}
\end{equation}
where $R_{\mathrm{UL}}$ is the total reserved UL PRBs and $U_{\mathrm{rsv}}$ is the portion of those PRBs unused by the victim.

We apply Reservation Waste carefully.
Under Prioritized with resource sharing, Waste must be read in the ATTACK window with active BSR on slice B, when the floor is semantically testable.
In the measured range at 106~PRB, slice B UL NPRB share in the ATTACK window falls from ${\approx}21.5\%$ under unprotected soft isolation (B0) to ${\approx}14\%$ with active MinRatio (B20--B80, Tab.~\ref{tab:frontier-sweep}), while $\delta$ p99 returns to the order of tens of ms and UAV UL returns to ${\approx}12.5$~Mbps (Tab.~\ref{tab:mitigations-map}).
Under MinRatio the victim recovers freshness and throughput: UL NPRB share falls because contention abates and fewer grant opportunities suffice to clear the load, since share indexes the fraction of scheduler opportunities in the window rather than a loss of capacity.
ReservationCost in the same window remains negative (${\approx}{-}0.71$), that is, $T_{\mathrm{cell}}$ under MinRatio stays above paper-stack B0, so the measured range keeps aggregate cell throughput at or above B0 for this protection, while ReservationWaste rises with the floor (from $0.50$ at B20 to $0.87$ at B80), making the cost of isolation visible at high floors.

\subsection{Control Map}

Table~\ref{tab:mitigations-map} maps controls under attack $N{=}4$ with video feedback at 12~Mbps (median of per-run values across 3 runs, mitigation fork).
AQM leaves $\delta$ near $12$~s, MinRatio restores tens of milliseconds, and rate limiting on slice A restores freshness from the attacker side, so the correct lever is an UL floor on the victim under the Network Resource Model rather than a local qdisc.

\begin{table}[t]
\centering
\caption{Control map under attack.}
\label{tab:mitigations-map}
\scriptsize
\setlength{\tabcolsep}{3pt}
\begin{tabular*}{\columnwidth}{@{\extracolsep{\fill}}lrrr@{}}
\toprule
Control & $\delta$ p99 & belief p99 & UAV UL \\
 &  & (m) & (Mbps) \\
\midrule
Paper stack (MinRatio B0) & $12.16{\pm}0.17$~s & $45.67{\pm}0.67$ & $8.04{\pm}0.06$ \\
AQM (\texttt{fq\_codel}) on host UL & $12.81{\pm}0.12$~s & $47.59{\pm}0.07$ & $7.71{\pm}0.10$ \\
Rate-limit on co-tenant (slice A) & $64.78{\pm}1.44$~ms & $1.07{\pm}0.00$ & $12.51{\pm}0.01$ \\
MinRatio UL B$=$20 & $57.88{\pm}2.62$~ms & $1.07{\pm}0.01$ & $12.50{\pm}0.02$ \\
MinRatio UL B$=$40 & $56.72{\pm}3.12$~ms & $1.07{\pm}0.01$ & $12.50{\pm}0.02$ \\
MinRatio UL B$=$60 & $56.11{\pm}4.81$~ms & $1.07{\pm}0.00$ & $12.49{\pm}0.02$ \\
MinRatio UL B$=$80 & $56.52{\pm}1.50$~ms & $1.07{\pm}0.01$ & $12.51{\pm}0.00$ \\
\bottomrule
\end{tabular*}
\end{table}


AQM on the UAV host (\texttt{fq\_codel}) keeps delivery age $\delta$ p99 at $12.81$~s and belief p99 at $47.59$~m.
RLC/PDCP/LCG buffers remain present in the final stage of this chain (Sec.~\ref{sec:lcg-hol})~\cite{3gpp_ts_38300,3gpp_ts_38321}.
Rate-limit on the adversarial co-tenant, although not standardized by 3GPP, reduces dose and restores AoI/belief near baseline, but acts on the attacking UE, whereas the uplink floor acts per S-NSSAI.
The uplink floor on the victim slice restores freshness with normative alignment~\cite{3gpp_ts_28541}.

UL MinRatio on slice B (values B$=$20/40/60/80 in the mitigation fork) restores delivery age $\delta$ p99 to the order of $56$ to $58$~ms and belief p99 to $1.07$~m, with UAV UL back to about $12.5$~Mbps (Tab.~\ref{tab:mitigations-map}).
The benefit of increasing the protection floor saturates early.
B$=$20 already delivers $\delta$ p99 of $57.88{\pm}2.62$~ms and UAV UL of $12.50{\pm}0.02$~Mbps, while raising the floor up to B$=$80 keeps $\delta$ p99 at $56.11$--$57.88$~ms and victim throughput at ${\approx}12.5$~Mbps.
In the measured range, MinRatio on the victim restores freshness with stable UAV UL, and the isolation$\times$efficiency curve is anchored by this early saturation of protection (Tab.~\ref{tab:frontier-sweep}).
This result identifies the lever aligned with the Network Resource Model, \texttt{rRMPolicyMinRatio} per S-NSSAI on the victim slice~\cite{3gpp_ts_28541}.
The floor under test in these artifacts is MinRatio on the UAS slice, while host AQM leaves intact the need for this floor.

On the standard testbed (B0, UL PF cell-wide), the same attack remains relevant ($\delta$ p99 ${\approx}12.2$~s).
AQM keeps $\delta$ p99 at ${\approx}12.8$~s, while restoration comes from UL MinRatio on the victim.
Rate-limit on the co-tenant also mitigates the effect, but acts on slice A.
Restoring freshness by victim isolation requires the full mitigation fork (UL RRMPolicy patch, dedicated startup, and flag).
The lever is normative, but structural.
It corresponds to \texttt{rRMPolicyMinRatio} of TS~28.541, but takes effect only with the mitigation fork, and reserving UL PRB reduces statistical multiplexing of the cell.
Testing the attack under an already active partial UL policy would change the experimental condition itself, since the uplink floor constitutes the defensive mechanism evaluated.
In artifacts B20--B80, this floor restores $\delta$, while paper stack B0 remains vulnerable.
The operational lesson is that direction, policy side, and UL enforcement matter as much as the value written in configuration.

\subsection{Frontier Isolation $\times$ Freshness $\times$ Efficiency}

Table~\ref{tab:frontier-sweep} structures the measured frontier at 106~PRB under attack $N{=}4$ and video feedback at 12~Mbps.
Floors B20 to B80 restore freshness while ReservationCost and ReservationWaste quantify multiplexing cost, so effective isolation is a measured tradeoff rather than a free configuration toggle.
Eqs.~\eqref{eq:reservation-cost}--\eqref{eq:reservation-waste} anchor Cost and Waste on this same curve.

In the B20--B80 range, protection saturates early: $\delta$ and UAV UL remain stable while ReservationWaste rises from $0.50$ to $0.87$.
ReservationCost stays at ${\approx}{-}0.71$ at all these floors, with aggregate $T_{\mathrm{cell}}$ statistically unchanged relative to B0.
At 51~PRB, the same B20--B80 floors restore $\delta$ p99 to $52.6$--$55.2$~ms with UAV UL ${\approx}12.6$--$12.7$~Mbps and Waste from $0.18$ to $0.80$, confirming the same qualitative direction under lower bandwidth.

\begin{table}[b]
\centering
\caption{MinRatio frontier on the victim slice.}
\label{tab:frontier-sweep}
\scriptsize
\setlength{\tabcolsep}{2.5pt}
\begin{tabular*}{\columnwidth}{@{\extracolsep{\fill}}lrrrrr@{}}
\toprule
Floor B & $\delta$ p99 & belief p99 & UAV UL & Share B & Waste \\
 &  & (m) & (Mbps) & UL (\%) &  \\
\midrule
B0 (paper stack) & $12.16{\pm}0.17$~s & $45.67{\pm}0.67$ & $8.04{\pm}0.06$ & ${\approx}21.5$ & n/a \\
B20 & $57.88{\pm}2.62$~ms & $1.07{\pm}0.01$ & $12.50{\pm}0.02$ & ${\approx}14.6$ & $0.50$ \\
B40 & $56.72{\pm}3.12$~ms & $1.07{\pm}0.01$ & $12.50{\pm}0.02$ & ${\approx}13.9$ & $0.74$ \\
B60 & $56.11{\pm}4.81$~ms & $1.07{\pm}0.00$ & $12.49{\pm}0.02$ & ${\approx}13.8$ & $0.83$ \\
B80 & $56.52{\pm}1.50$~ms & $1.07{\pm}0.01$ & $12.51{\pm}0.00$ & ${\approx}14.3$ & $0.87$ \\
\bottomrule
\end{tabular*}
\end{table}

\begin{table*}[!t]
\centering
\caption{Comparison with related work.}
\label{tab:related-comparison}
\footnotesize
\setlength{\tabcolsep}{4.5pt}
\begin{tabular*}{\textwidth}{@{\extracolsep{\fill}}lcccccc@{}}
\toprule
Work & Cross-slice & UL axis & Weak adv. & Freshness & Cyber-phys. & Ctrl.\ map \\
\midrule
Baguer et al.~\cite{baguer_enabling_2024} & \fbFull & \fbHalf & \fbEmpty & \fbHalf & \fbFull & \fbHalf \\
Moreira et al.~\cite{moreira_noisy_neighbor_influence_2026} & \fbFull & \fbEmpty & \fbHalf & \fbEmpty & \fbEmpty & \fbHalf \\
Zipper~\cite{balasingam_application_level_service_2024} & \fbFull & \fbHalf & \fbEmpty & \fbEmpty & \fbEmpty & \fbFull \\
Xu et al.~\cite{xu_integrity_under_siege_2025} & \fbFull & \fbEmpty & \fbEmpty & \fbEmpty & \fbEmpty & \fbHalf \\
DoSQ~\cite{ashik_dosq_cross_layer_2026} & \fbEmpty & \fbEmpty & \fbEmpty & \fbEmpty & \fbEmpty & \fbHalf \\
Huang et al.~\cite{huang_gray_failure_2017} & \fbEmpty & \fbEmpty & \fbEmpty & \fbHalf & \fbEmpty & \fbEmpty \\
WiSwarm~\cite{tripathi_wiswarm_age_information_2022} & \fbEmpty & \fbHalf & \fbEmpty & \fbFull & \fbFull & \fbEmpty \\
\midrule
\textbf{This work} & \fbFull & \fbFull & \fbFull & \fbFull & \fbFull & \fbFull \\
\bottomrule
\end{tabular*}
\end{table*}

\section{Discussion and Threats to Validity}
\label{sec:discussion}

Results must be interpreted considering the boundary of the stack used in the experiments.
They belong to a single reproducible implementation based on OAI~\cite{openairinterface} and ORANSlice~\cite{cheng_oranslice_open_source_2024}, and characterize the class of deployments already delimited in Setup (Sec.~\ref{sec:setup}), with soft isolation and incomplete or asymmetric UL enforcement.
Still, OAI, ORANSlice, and ArduPilot SITL represent a realistic class of 5G SA implementations with RAN slicing and a closed C2 loop, already used in O-RAN testbeds and published UAV evaluations~\cite{cheng_oranslice_open_source_2024,villa_x5g_open_programmable_2025,baldi_ardupilot_based_adaptive_2022,ardupilot_sitl}, and not an ad-hoc harness.

The methodological value of rfsim lies in removing fading, fine radio mobility, and external interference to isolate the UL scheduler as the causal variable under controlled data-plane emulation.
The mechanism under study is scheduling-layer grant distribution on the UL, above the radio channel that rfsim abstracts~\cite{oai_rfsimulator,rouili_evaluating_open_source_2024}.
Silent State Staleness emerges when authorized neighbor demand depletes victim grant opportunities under soft isolation, a deficit that the PRE/ATTACK/DRAIN signature and the dose and video controls attribute to the scheduler (Sec.~\ref{sec:eval}).
Fading and mobility would modulate the aging magnitude while leaving the grant-deficit origin intact.
OTA corroboration on commercial radio would strengthen external confidence and remains future work.
The \texttt{max\_sched\_ues} sweep leaves $\delta$ near $12.6$~s across caps, confirming that headline magnitude tracks UL contention rather than the scheduler UE cap.
What generalizes is the possibility that a work-conserving~\cite{johnson_nexran_closed_loop_2021} implementation creates a surface in which authorized neighbor demand ages freshness-sensitive updates.

The video feedback load used here is synthetic and controlled.
It acts as a load conditioner that brings the victim UE close to a plausible temporal boundary in BVLOS operations with auxiliary video, at the rate-matched granularity of this campaign.
Real encoders with adaptive bitrate would alter the instantaneous pattern~\cite{baltaci_analyzing_real_time_2022}.
Likewise, SITL makes observable the separation between the vehicle, which can remain intact, and the GCS, which begins to operate on outdated state.

The notion of stealth in this work is deliberately restricted.
We assert false-healthy as a verifiable predicate (Eq.~\eqref{eq:sss}), in which the GCS--UAV link remains operational (GUIDED, inactive \texttt{failsafe}, heartbeats still delivered) while state freshness at the GCS collapses.
We use ``silent'' in the sense of invisible to conventional connectivity indicators~\cite{huang_gray_failure_2017}.
The corresponding phenomenon is Silent State Staleness.
The opposite-leg OWD/IPDV/availability probe matches the conventional monitor the operator observes, while the contrast with freshness on the telemetry leg constitutes false-healthy (Sec.~\ref{sec:paths}).
Instrumenting the UAV$\rightarrow$GCS leg in parallel for OWD and IPDV remains complementary measurement that would remove residual ambiguity on path KPIs.
MAVLink heartbeat gaps (median of 32 gaps ${>}$1.5~s, max gap ${\approx}$2.9--3.2~s under $N{=}4$) are secondary observables.
A freshness-aware monitor based on delivery age would detect the phenomenon, turning this observation into a defense opportunity~\cite{yates_age_information_introduction_2021}.

FlyBlind establishes an error-enabling condition in situational awareness, whose operational unfolding depends on mission window, operator policy, presence of video feedback, and adversarial dose~\cite{sorbelli_risk_2024,tripathi_wiswarm_age_information_2022}.
Spatial impact scales with airspeed and orbit geometry, so relating $\delta_i$ to operational envelopes is a natural extension of the belief-vs-truth reading.
Beyond the UL floor per S-NSSAI already mapped (Sec.~\ref{sec:mitigations}), operators may adopt intra-UE mitigations within the same subscriber boundary, including distinct 5QI or LCG mappings for MAVLink versus video and logical channel prioritization~\cite{3gpp_ts_23501,3gpp_ts_38321}, as complementary defense mechanisms beyond this campaign.

\section{Related Work}
\label{sec:related}

We situate FlyBlind relative to nearby work on UAS/5G, noisy neighbor, slicing assurance, and classic DoS.
Table~\ref{tab:related-comparison} places FlyBlind on the combination of attacked freshness, weak authorized adversary, false-healthy observability, and control map, relative to nearby work rather than as a single matrix cell.
Full, half-filled, and empty circles mark presence, partial coverage, and absence on each axis.

Baguer et al.\ study UAS BVLOS over O-RAN in a benign scenario and point to the need for a pilot-side divergence metric left open by that work~\cite{baguer_enabling_2024}.
FlyBlind occupies the adversarial complement on the UL axis: an authorized co-tenant induces Silent State Staleness under soft isolation, with delivery age and \texttt{belief\_vs\_truth} while conventional KPIs remain green~\cite{elayoubi_5g_ran_slicing_2019,vittal_preventing_cross_network_2023}.

On the core plane, contention between slices has already been characterized as noisy neighbor at the UPF by Moreira et al.~\cite{moreira_noisy_neighbor_influence_2026}.
Our focus is before that point in the pipeline, on distribution of UL opportunities between slices at the RAN, and not on core GTP-U decapsulation~\cite{zhang_invade_2025}.

The distance between slice policy and observed guarantee also appears on the assurance side.
Zipper shows that RAN slicing optimized for slice-level assurance leaves application-level service assurance as a separate problem~\cite{balasingam_application_level_service_2024}.
Building on this intuition, we replace the service assurance objective with an operational security reading oriented to freshness~\cite{nassi_sok_2021,huang_fira_2026}.
Relative to RIC/xApp lines that allocate PRBs across slices to meet SLA objectives such as latency and throughput~\cite{dorcheh_dora_dynamic_o_2025}, FlyBlind supplies a concrete operator need: enforce UL floors for timeliness-sensitive tenants under soft isolation.

In resource attacks, the classic DoS/DDoS taxonomy describes attacks that exhaust a resource until they prevent legitimate use of the service~\cite{mirkovic_taxonomy_ddos_attack_2004}.
FlyBlind preserves apparent availability and attacks state freshness, characterizing a failure mode distinct from disconnection.

On the radio side, Xu et al.\ and DoSQ study attacks with stronger adversaries, such as rogue gNB, side channels, or jamming~\cite{xu_integrity_under_siege_2025,ashik_dosq_cross_layer_2026}.
We differ by showing that an adversary restricted to authorized UEs and legitimate UL traffic already induces relevant operational failure when cross-slice UL isolation is insufficient.

Huang et al.\ present gray failure and differential observability~\cite{huang_gray_failure_2017}.
Xing et al.\ show that infrastructural details below the application can generate security consequences when monitoring concentrates on the wrong property~\cite{xing_criticality_2024}.
We bring these ideas to UAS over 5G and formalize a false-healthy state in which IP health and freshness health separate.

Surveys of cellular communications for UAV and 5G/6G development agendas identify command and control (C2) connectivity as a critical requirement~\cite{geraci_what_2022,fotouhi_survey_uav_cellular_2019,abdalla_communications_standards_unmanned_2021}.
Yates et al.\ and WiSwarm position AoI as a natural metric for updates and control~\cite{yates_age_information_introduction_2021,tripathi_wiswarm_age_information_2022}.
FlyBlind uses AoI as a tool to expose a security surface in 5G SA for BVLOS~\cite{sorbelli_risk_2024}, orthogonal to AoI theory, noisy neighbor at the core, and radio attacks with a strong adversary~\cite{yates_age_information_introduction_2021,moreira_noisy_neighbor_influence_2026,xu_integrity_under_siege_2025,ashik_dosq_cross_layer_2026}.
To the best of our knowledge, FlyBlind is the first to combine state freshness as an attacked property, an authorized adversary without a rogue gNB or jamming, a falsifiable false-healthy predicate, and cyber-physical evidence under UL cross-slice contention.

\section{Conclusion}
\label{sec:conclusion}

Recent work on slice isolation and UAS C2 monitoring over 5G networks measures mostly link health and may overestimate operational security of BVLOS UAS when the critical property is state freshness at the GCS. We present FlyBlind to test the false-healthy predicate using measurable criteria under an authorized adversary and demonstrate that, with soft isolation and realistic video feedback, a co-tenant induces Silent State Staleness with measurable divergence in GCS belief, while monitored connectivity remains operational and the vehicle remains intact. We conclude that deployments of this class need effective uplink floors per S-NSSAI and explicit freshness monitoring, in addition to reachability, OWD, IPDV, availability, and link integrity, at a reservation cost that operators must budget for. The false-healthy phenomenon we characterize is an adversarial and cyber-physical instance of gray failure, induced by authorized co-tenancy and not by accidental datacenter failure or a powerful adversary. This reinforces the pilot-side perspective as a security surface. We hope the study and artifacts encourage more rigorous evaluations of C2/UAS over 5G.

\bibliographystyle{plainurl}
\bibliography{References}

@article{pradosgarzon_5g_non_public_2021,
    author = {Prados-Garzon, Jonathan and Ameigeiras, Pablo and Ordonez-Lucena, Jose and Munoz, Pablo and Adamuz-Hinojosa, Oscar and Camps-Mur, Daniel},
    title = {{5G} Non-Public Networks: Standardization, Architectures and Challenges},
    journal = {{IEEE} Access},
    volume = {9},
    year = {2021},
    pages = {153893--153908}
}

@article{elayoubi_5g_ran_slicing_2019,
    author = {Elayoubi, Salah Eddine and Jemaa, Sana Ben and Altman, Zwi and Galindo-Serrano, Ana},
    title = {{5G} {RAN} Slicing for Verticals: Enablers and Challenges},
    journal = {{IEEE} Communications Magazine},
    volume = {57},
    number = {1},
    year = {2019},
    month = {Jan},
    pages = {28--34}
}

@techreport{5gacia_npn_industrial_2019,
    author = {{5G-ACIA}},
    title = {{5G} Non-Public Networks for Industrial Scenarios},
    institution = {{5G} Alliance for Connected Industries and Automation},
    year = {2019},
    note = {White paper. Accessed: 2026-08-19}
}

@article{mirkovic_taxonomy_ddos_attack_2004,
    author = {Mirkovic, Jelena and Reiher, Peter},
    title = {A taxonomy of {DDoS} attack and {DDoS} defense mechanisms},
    journal = {{ACM} {SIGCOMM} Computer Communication Review},
    volume = {34},
    number = {2},
    year = {2004},
    month = {Apr},
    pages = {39--53}
}

@article{yates_age_information_introduction_2021,
    author = {Yates, Roy D. and Sun, Yin and Brown, D. Richard and Kaul, Sanjit K. and Modiano, Eytan and Ulukus, Sennur},
    title = {Age of Information: An Introduction and Survey},
    journal = {{IEEE} Journal on Selected Areas in Communications},
    volume = {39},
    number = {5},
    year = {2021},
    month = {May},
    pages = {1183--1210}
}

@inproceedings{baltaci_analyzing_real_time_2022,
    author = {Baltaci, Aygün and Cech, Hendrik and Mohan, Nitinder and Geyer, Fabien and Bajpai, Vaibhav and Ott, Jörg and Schupke, Dominic},
    title = {Analyzing real-time video delivery over cellular networks for remote piloting aerial vehicles},
    booktitle = {Proceedings of the 22nd {ACM} Internet Measurement Conference},
    publisher = {Association for Computing Machinery},
    address = {New York, NY, USA},
    year = {2022},
    month = {Oct},
    pages = {98--112}
}

@inproceedings{balasingam_application_level_service_2024,
    author = {Balasingam, Arjun and Kotaru, Manikanta and Bahl, Paramvir},
    title = {Application-Level Service Assurance with {5G} {RAN} Slicing},
    booktitle = {{USENIX} Symposium on Networked Systems Design and Implementation ({NSDI})},
    year = {2024},
    pages = {841--857}
}

@article{baldi_ardupilot_based_adaptive_2022,
    author = {Baldi, Simone and Sun, Danping and Xia, Xin and Zhou, Guopeng and Liu, Di},
    title = {ArduPilot-Based Adaptive Autopilot: Architecture and Software-in-The-Loop Experiments},
    journal = {{IEEE} Transactions on Aerospace and Electronic Systems},
    volume = {58},
    number = {5},
    year = {2022},
    pages = {4473--4485}
}

@article{staff_bufferbloat_what_s_2012,
    author = {Staff, CACM},
    title = {BufferBloat: what's wrong with the internet?},
    journal = {Communications of the {ACM}},
    volume = {55},
    number = {2},
    year = {2012},
    month = {Feb},
    pages = {40--47}
}

@article{abdalla_communications_standards_unmanned_2021,
    author = {Abdalla, Aly Sabri and Marojevic, Vuk},
    title = {Communications Standards for Unmanned Aircraft Systems: The {3GPP} Perspective and Research Drivers},
    journal = {{IEEE} Communications Standards Magazine},
    volume = {5},
    number = {1},
    year = {2021},
    month = {Mar},
    pages = {70--77}
}

@misc{dorcheh_dora_dynamic_o_2025,
    author = {Dorcheh, Alireza Ebrahimi and Seyfi, Tolunay and Afghah, Fatemeh},
    title = {{DORA}: Dynamic O-{RAN} Resource Allocation for Multi-Slice {5G} Networks},
    howpublished = {arXiv:2509.07242},
    note = {preprint},
    year = {2025},
    month = {Sep}
}

@misc{ashik_dosq_cross_layer_2026,
    author = {Ashik, Mahmudul Hassan and Hossain, Moinul},
    title = {{DoSQ}: A Cross-Layer Denial of Service Quality Attack by Exploiting Side Channels in {5G} {NR}},
    howpublished = {arXiv:2607.16102},
    note = {preprint},
    year = {2026},
    month = {Jul}
}

@inproceedings{guo_dynamic_tdd_interference_2018,
    author = {Guo, Shaozhen and Hou, Xiaolin and Wang, Hanning},
    title = {Dynamic {TDD} and interference management towards {5G}},
    booktitle = {2018 {IEEE} Wireless Communications and Networking Conference ({WCNC})},
    publisher = {IEEE},
    year = {2018},
    month = {Apr},
    pages = {1--6}
}

@article{baguer_enabling_2024,
    author = {Baguer, Pau and Municio, Esteban and Garcia-Aviles, Gines and Costa-Pérez, Xavier},
    title = {Enabling Beyond-Visual-Line-of-Sight Drones Operation Over Open {RAN} {5G} Networks With Slicing},
    journal = {{IEEE} Network},
    volume = {38},
    number = {6},
    pages = {163--169},
    year = {2024},
    month = {Nov}
}

@inproceedings{rouili_evaluating_open_source_2024,
    author = {Rouili, Mohamed and Saha, Niloy and Golkarifard, Morteza and Zangooei, Mohammad and Boutaba, Raouf and Onur, Ertan and Saleh, Aladdin},
    title = {Evaluating Open-Source {5G} {SA} Testbeds: Unveiling Performance Disparities in {RAN} Scenarios},
    booktitle = {{NOMS} 2024-2024 {IEEE} Network Operations and Management Symposium},
    year = {2024},
    month = {May},
    pages = {1--6}
}

@inproceedings{huang_fira_2026,
    author = {Huang, Yizhi and Oygenblik, David and Zhang, Runze and Yao, Mingxuan and Ibrahim, Muhammad and Sahin, Burak and Xu, Haichuan and Zonouz, Saman and Saltaformaggio, Brendan},
    title = {{FIRA}: Enabling Automatic Forensic Investigation of Unmanned Aerial Vehicles},
    booktitle = {35th {USENIX} Security Symposium ({USENIX} Security 26)},
    year = {2026}
}

@article{fotouhi_survey_uav_cellular_2019,
    author = {Fotouhi, Azade and Qiang, Haoran and Ding, Ming and Hassan, Mahbub and Giordano, Lorenzo Galati and Garcia-Rodriguez, Adrian and Yuan, Jinhong},
    title = {Survey on {UAV} Cellular Communications: Practical Aspects, Standardization Advancements, Regulation, and Security Challenges},
    journal = {{IEEE} Communications Surveys \& Tutorials},
    volume = {21},
    number = {4},
    year = {2019},
    pages = {3417--3442}
}

@article{villa_x5g_open_programmable_2025,
    author = {Villa, Davide and Khan, Imran and Kaltenberger, Florian and Hedberg, Nicholas and da Silva, Rúben Soares and Maxenti, Stefano and Bonati, Leonardo and Kelkar, Anupa and Dick, Chris and Baena, Eduardo and Jornet, Josep M. and Melodia, Tommaso and Polese, Michele and Koutsonikolas, Dimitrios},
    title = {{X5G}: An Open, Programmable, Multi-Vendor, End-to-End, Private {5G} O-{RAN} Testbed With {NVIDIA} {ARC} and OpenAirInterface},
    journal = {{IEEE} Transactions on Mobile Computing},
    volume = {24},
    number = {11},
    year = {2025},
    month = {Nov},
    pages = {11305--11322}
}

@misc{xu_integrity_under_siege_2025,
    author = {Xu, Jiali and Loscri, Valeria and Rouvoy, Romain},
    title = {Integrity Under Siege: A Rogue gNodeB's Manipulation of {5G} Network Slice Allocation},
    howpublished = {arXiv:2511.03312},
    note = {preprint},
    year = {2025},
    month = {Nov}
}

@inproceedings{zhang_invade_2025,
    author = {Zhang, Yiming and Wan, Tao and Yang, Yaru and Duan, Haixin and Wang, Yichen and Chen, Jianjun and Wei, Zixiang and Li, Xiang},
    title = {Invade the Walled Garden: Evaluating {GTP} Security in Cellular Networks},
    booktitle = {2025 {IEEE} Symposium on Security and Privacy ({SP})},
    year = {2025},
    month = {May},
    pages = {1159--1177}
}

@inproceedings{allouch_mavsec_securing_mavlink_2019,
    author = {Allouch, Azza and Cheikhrouhou, Omar and Koubâa, Anis and Khalgui, Mohamed and Abbes, Tarek},
    title = {MAVSec: Securing the {MAVLink} Protocol for Ardupilot/{PX4} Unmanned Aerial Systems},
    booktitle = {2019 15th International Wireless Communications \& Mobile Computing Conference ({IWCMC})},
    year = {2019},
    month = {Jun},
    pages = {621--628}
}

@inproceedings{johnson_nexran_closed_loop_2021,
    author = {Johnson, David and Maas, Dustin and Van Der Merwe, Jacobus},
    title = {NexRAN: Closed-loop {RAN} slicing in {POWDER} -A top-to-bottom open-source open-{RAN} use case},
    booktitle = {Proceedings of the 15th {ACM} Workshop on Wireless Network Testbeds, Experimental evaluation \& CHaracterization},
    publisher = {Association for Computing Machinery},
    address = {New York, NY, USA},
    year = {2022},
    pages = {17--23}
}

@misc{moreira_noisy_neighbor_influence_2026,
    author = {Moreira, Rodrigo and Moreira, Larissa Ferreira Rodrigues and Carvalho, Tereza C. and Silva, Flavio de Oliveira},
    title = {Noisy Neighbor Influence in the Data Plane of Beyond {5G} Networks},
    howpublished = {arXiv:2601.12106},
    note = {preprint},
    year = {2026},
    month = {Jan}
}

@inproceedings{xing_criticality_2024,
    author = {Xing, Jiarong and Yoo, Sophia and Foukas, Xenofon and Kim, Daehyeok and Reiter, Michael K.},
    title = {On the Criticality of Integrity Protection in {5G} Fronthaul Networks},
    booktitle = {33rd {USENIX} Security Symposium ({USENIX} Security 24)},
    year = {2024},
    pages = {4463--4479}
}

@inproceedings{cheng_oranslice_open_source_2024,
    author = {Cheng, Hai and D'Oro, Salvatore and Gangula, Rajeev and Velumani, Sakthivel and Villa, Davide and Bonati, Leonardo and Polese, Michele and Melodia, Tommaso and Arrobo, Gabriel and Maciocco, Christian},
    title = {{ORANSlice}: An Open Source {5G} Network Slicing Platform for O-{RAN}},
    booktitle = {Proceedings of the 30th Annual International Conference on Mobile Computing and Networking},
    publisher = {Association for Computing Machinery},
    address = {New York, NY, USA},
    year = {2024},
    month = {Dec},
    pages = {2297--2302}
}

@inproceedings{vittal_preventing_cross_network_2023,
    author = {Vittal, Shwetha and Dixit, Unnati and Sovitkar, Siddhesh Pratim and Sowjanya, K and Antony Franklin, A},
    title = {Preventing Cross Network Slice Disruptions in a Zero-Trust and Multi-Tenant Future {5G} Networks},
    booktitle = {2023 {IEEE} 9th International Conference on Network Softwarization ({NetSoft})},
    year = {2023},
    month = {Jun},
    pages = {227--231}
}

@article{wen_private_2022,
    author = {Wen, Miaowen and Li, Qiang and Kim, Kyeong Jin and López-Pérez, David and Dobre, Octavia A. and Poor, H. Vincent and Popovski, Petar and Tsiftsis, Theodoros A.},
    title = {Private {5G} Networks: Concepts, Architectures, and Research Landscape},
    journal = {{IEEE} Journal of Selected Topics in Signal Processing},
    volume = {16},
    number = {1},
    pages = {7--25},
    year = {2022},
    month = {Jan}
}

@inproceedings{kaul_real_time_status_2012,
    author = {Kaul, Sanjit and Yates, Roy and Gruteser, Marco},
    title = {Real-time status: How often should one update?},
    booktitle = {{IEEE} {INFOCOM}},
    publisher = {IEEE},
    year = {2012},
    month = {Mar},
    pages = {2731--2735}
}

@misc{konapalli_reverse_engineering_control_2025,
    author = {Konapalli, Yasaswini and Ben Othmane, Lotfi and Tunc, Cihan and Benchellal, Feras and Mudagere, Likhita},
    title = {Reverse Engineering and Control-Aware Security Analysis of the {ArduPilot} {UAV} Framework},
    year = {2025},
    howpublished = {arXiv:2512.01164},
    note = {arXiv preprint}
}

@inproceedings{sorbelli_risk_2024,
    author = {Sorbelli, Francesco Betti and Chatterjee, Punyasha and Das, Papiya and Pinotti, Cristina M.},
    title = {Risk Assessment in {BVLoS} Operations for {UAVs}: Challenges and Solutions},
    booktitle = {2024 20th International Conference on Distributed Computing in Smart Systems and the Internet of Things ({DCOSS-IoT})},
    year = {2024},
    month = {Apr},
    pages = {300--307}
}

@inproceedings{nassi_sok_2021,
    author = {Nassi, Ben and Bitton, Ron and Masuoka, Ryusuke and Shabtai, Asaf and Elovici, Yuval},
    title = {{SoK}: Security and Privacy in the Age of Commercial Drones},
    booktitle = {2021 {IEEE} Symposium on Security and Privacy ({SP})},
    year = {2021},
    month = {May},
    pages = {1434--1451}
}

@inproceedings{akundi_suppressing_noisy_neighbours_2020,
    author = {Akundi, Salil and Prabhu, Shailesh and Upadhya, Nithin B. K. and Mondal, Subhas Chandra},
    title = {Suppressing Noisy Neighbours in 5G networks: An end-to-end NFV-based framework to detect and suppress noisy neighbours},
    booktitle = {Proceedings of the 21st International Conference on Distributed Computing and Networking ({ICDCN})},
    publisher = {Association for Computing Machinery},
    year = {2020}
}

@techreport{3gpp_ts_22125,
    author = {{3GPP}},
    title = {{TS 22.125: Unmanned Aerial System (UAS) support in 3GPP (Release 17)}},
    year = {2022},
    institution = {3rd Generation Partnership Project (3GPP)}
}

@techreport{3gpp_ts_23501,
    author = {{3GPP}},
    title = {{TS 23.501: System architecture for the 5G System (5GS) (Release 17)}},
    institution = {3rd Generation Partnership Project (3GPP)},
    year = {2022}
}

@techreport{3gpp_ts_28541,
    author = {{3GPP}},
    title = {{TS 28.541: Management and orchestration; 5G Network Resource Model (NRM); Stage 2 and stage 3 (Release 17)}},
    institution = {3rd Generation Partnership Project (3GPP)},
    year = {2023},
    url = {https://www.3gpp.org/DynaReport/28541.htm},
    note = {ETSI TS 128 541 V17.10.0 (2023-05)}
}

@techreport{3gpp_ts_28554,
    author = {{3GPP}},
    title = {{TS 28.554: Management and orchestration; 5G end to end Key Performance Indicators (KPI) (Release 17)}},
    institution = {3rd Generation Partnership Project (3GPP)},
    year = {2023},
    url = {https://www.3gpp.org/DynaReport/28554.htm},
    note = {ETSI TS 128 554 V17.10.0 (2023-07)}
}

@techreport{3gpp_ts_38300,
    author = {{3GPP}},
    title = {{TS 38.300: NR; NR and NG-RAN Overall description; Stage-2 (Release 17)}},
    institution = {3rd Generation Partnership Project (3GPP)},
    year = {2022},
    url = {https://www.3gpp.org/DynaReport/38300.htm}
}

@techreport{3gpp_ts_38321,
    author = {{3GPP}},
    title = {{TS 38.321: NR; Medium Access Control (MAC) protocol specification (Release 17)}},
    institution = {3rd Generation Partnership Project (3GPP)},
    year = {2023},
    url = {https://www.3gpp.org/DynaReport/38321.htm},
    note = {ETSI TS 138 321 V17.6.0 (2023-10)}
}

@techreport{3gpp_ts_38101,
    author = {{3GPP}},
    title = {{TS 38.101-1: NR; User Equipment (UE) radio transmission and reception; Part 1: Range 1 Standalone (Release 17)}},
    institution = {3rd Generation Partnership Project (3GPP)},
    year = {2022},
    url = {https://www.3gpp.org/DynaReport/38101-1.htm},
    note = {ETSI TS 138 101-1 V17.6.0 (2022-08)}
}

@inproceedings{jiang_understanding_bufferbloat_cellular_2012,
    author = {Jiang, Haiqing and Liu, Zeyu and Wang, Yaogong and Lee, Kyunghan and Rhee, Injong},
    title = {Understanding bufferbloat in cellular networks},
    booktitle = {Proceedings of the 2012 {ACM} {SIGCOMM} workshop on Cellular networks: operations, challenges, and future design},
    publisher = {Association for Computing Machinery},
    address = {New York, NY, USA},
    year = {2012},
    month = {Aug},
    pages = {1--6}
}

@misc{demichelis_ip_packet_delay_variation_2002,
    author = {Demichelis, C. and Chimento, P.},
    title = {{IP Packet Delay Variation Metric} for {IP Performance Metrics} ({IPPM})},
    howpublished = {{RFC} 3393},
    year = {2002},
    month = {Nov},
    url = {https://www.rfc-editor.org/rfc/rfc3393.html},
    note = {Standards Track. Accessed: 2026-08-24}
}

@techreport{honeywelltestreport263do377c2linktest_us_2023,
    author      = {{Honeywell}},
    title       = {{FAA BAA Call 3: UAS Command \& Control (006): Final Test Report (FTR), Full Report with Appendices B \& C}},
    institution = {Honeywell},
    year        = {2023},
    month       = nov,
    note        = {Revision 2.0},
    url         = {https://www.faa.gov/uas/programs_partnerships/test_sites/TestReport-263_Honeywell_20231120-full-Rev2.pdf}
}

@techreport{mitllad1197090c2linksrmanalysisdo377a_us_2022,
    author = {Serres, Christine and Gill, Bilal and Reheis, Peter and Edwards, Matthew},
    title = {{RTCA} Detect and Avoid Phase 2: Safety Risk Management Modeling and Simulation Final Report},
    institution = {{MIT} Lincoln Laboratory},
    year = {2022},
    url = {https://doi.org/10.5281/zenodo.7778187},
    note = {Accession {AD}1197090; supports {DO-377} {C2} {MASPS}. Accessed: 2026-08-19}
}

@article{geraci_what_2022,
    author = {Geraci, Giovanni and Garcia-Rodriguez, Adrian and Azari, M. Mahdi and Lozano, Angel and Mezzavilla, Marco and Chatzinotas, Symeon and Chen, Yun and Rangan, Sundeep and Renzo, Marco Di},
    title = {What Will the Future of {UAV} Cellular Communications Be? A Flight From {5G} to {6G}},
    journal = {{IEEE} Communications Surveys \& Tutorials},
    volume = {24},
    number = {3},
    pages = {1304--1335},
    year = {2022}
}

@misc{tripathi_wiswarm_age_information_2022,
    author = {Tripathi, Vishrant and Kadota, Igor and Tal, Ezra and Rahman, Muhammad Shahir and Warren, Alexander and Karaman, Sertac and Modiano, Eytan},
    title = {{WiSwarm}: Age-of-Information-based Wireless Networking for Collaborative Teams of {UAVs}},
    howpublished = {arXiv:2212.03298},
    note = {preprint},
    year = {2022},
    month = {Dec}
}

@misc{kubernetes,
    author = {{Cloud Native Computing Foundation}},
    title = {Kubernetes},
    year = {2026},
    url = {https://kubernetes.io/},
    note = {Accessed: 2026-05-06}
}

@misc{kind,
    author = {{The Kubernetes Authors}},
    title = {kind: Kubernetes {IN} Docker},
    year = {2026},
    url = {https://kind.sigs.k8s.io/},
    note = {Accessed: 2026-05-06}
}

@misc{pymavlink,
    author = {{ArduPilot}},
    title = {pymavlink: Python {MAVLink} Interface and Utilities},
    year = {2026},
    url = {https://github.com/ArduPilot/pymavlink},
    note = {Accessed: 2026-05-06}
}

@misc{mavlink_guide,
    author = {{MAVLink Development Team}},
    title = {{MAVLink} Guide},
    year = {2025},
    url = {https://mavlink.io/},
    note = {Accessed: 2026-05-06}
}

@inproceedings{huang_gray_failure_2017,
    author = {Huang, Peng and Guo, Chuanxiong and Zhou, Lidong and Lorch, Jacob R. and Dang, Yingnong and Chintalapati, Murali and Yao, Randolph},
    title = {Gray Failure: The {Achilles}' Heel of Cloud-Scale Systems},
    booktitle = {Proceedings of the 16th Workshop on Hot Topics in Operating Systems ({HotOS})},
    year = {2017},
    pages = {150--155}
}

@misc{openairinterface,
    author = {{OpenAirInterface Software Alliance}},
    title = {{OpenAirInterface}: Open Source Software for {5G} {RAN} and Core},
    year = {2026},
    url = {https://openairinterface.org/},
    note = {Accessed: 2026-08-21}
}

@misc{oai_rfsimulator,
    author = {{OpenAirInterface Software Alliance}},
    title = {{RF} Simulator ({rfsim}) Documentation},
    year = {2026},
    url = {https://github.com/OPENAIRINTERFACE/openairinterface5g/blob/develop/radio/rfsimulator/README.md},
    note = {Accessed: 2026-08-21}
}

@misc{ardupilot,
    author = {{ArduPilot Dev Team}},
    title = {{ArduPilot}: Open Source Autopilot},
    year = {2026},
    url = {https://ardupilot.org},
    note = {Accessed: 2026-08-21}
}

@misc{ardupilot_sitl,
    author = {{ArduPilot Dev Team}},
    title = {{SITL} Simulator (Software in the Loop)},
    year = {2026},
    url = {https://ardupilot.org/dev/docs/sitl-simulator-software-in-the-loop.html},
    note = {Accessed: 2026-08-21}
}

@misc{ardupilot_simulation,
    author = {{ArduPilot Dev Team}},
    title = {Simulation},
    year = {2026},
    url = {https://ardupilot.org/dev/docs/simulation-2.html},
    note = {Accessed: 2026-08-21}
}

@misc{iperf3,
    author = {{ESnet} and {Lawrence Berkeley National Laboratory}},
    title = {{iperf3}: A {TCP}, {UDP}, and {SCTP} Network Bandwidth Measurement Tool},
    year = {2026},
    url = {https://github.com/esnet/iperf},
    note = {Accessed: 2026-08-21}
}

\appendix

\section*{Ethical Considerations}
\label{app:ethics}

We ran all experimental work in an isolated laboratory under our administrative control.
No third-party traffic and no real aerial operation occurred.
Video feedback is synthetic (\texttt{iperf} rate-matched), and the cyber-physical loop is closed in ArduPilot SITL.
Adversarial UEs used only authorized laboratory SIMs.
The mechanism under study does not require compromising RAN, core, or MAVLink.
An authorized subscriber with legitimate UL demand on a neighbor slice suffices.
The study involved no human subjects and required no IRB review.

We acknowledge the dual-use character of this work.
A malicious reader could treat the artifacts as a recipe for UL contention across slices on a production cell.
If that recipe were applied to a live BVLOS link, an operator could act on stale GCS state while conventional link monitors remain green.
We mitigate this risk by releasing the work as a defensive instrument that helps operators harden UL soft isolation through \texttt{rRMPolicyMinRatio} floors and freshness monitoring ($A_{\mathrm{fresh}}$), rather than as a guide for attacking production networks.
We introduce no zero-day vulnerability that targets a specific vendor.
The phenomenon rests on public 3GPP policies and already documented open-source stacks, and we do not disclose proprietary vulnerabilities outside that scope.

\section*{Open Science}
\label{app:openscience}

To support reproducibility and follow-on work, we make the artifacts needed to evaluate our central claims available through an anonymized repository.
The repository includes Kind and Kubernetes scripts, OAI and ORANSlice configurations, ArduPilot SITL integration, frozen campaign outputs under \texttt{02\_results/}, analysis scripts, the mitigation fork, and documentation of the experimental environment.

The materials are organized so that a reader can follow the experimental pipeline: (i) bring up the sliced 5G SA testbed; (ii) run the UAV C2 loop under soft isolation; (iii) execute the FlyBlind dose, video, and bandwidth campaigns; and (iv) recompute the freshness, belief, and isolation metrics that support the false-healthy predicate.

The repository redistributes only what licenses allow.
Third-party components such as OAI, ORANSlice, and ArduPilot remain cited and are obtained from their upstream projects when redistribution is restricted.

The anonymized artifact repository is available at:
\begin{center}
\url{https://anonymous.4open.science/r/flyblind-94C6/README.md}
\end{center}
The link uses anonymous.4open.science with conference ID \texttt{SEC27} to preserve double-blind review.
Once the paper has been accepted, the repository will be made public.

\end{document}